\documentclass{article}
\usepackage{epsfig}
\usepackage[centertags]{amsmath}
\usepackage{graphicx}

\newcommand{\F}{\mathcal{F}}
\newcommand{\N}{\mathcal{N}}

\numberwithin{equation}{section}

\newcommand{\QED}{\hspace*{\fill}\rule{2.5mm}{2.5mm}}

\usepackage{color}
\newcommand\qed{\hfill$\sqcap\kern-7.5pt\hbox{$\sqcup$}$}

\newcommand{\kk}{{\bf k}}

\newcommand{\beqn}{\begin{equation}}
\newcommand{\eeqn}{\end{equation}}
\newcommand{\bear}{\begin{eqnarray}}
\newcommand{\eear}{\end{eqnarray}}
\newcommand{\bean}{\begin{eqnarray*}}
\newcommand{\eean}{\end{eqnarray*}}

\allowdisplaybreaks

\begin{document}

\title{A kinetic model for very low temperature dilute Bose gases}

\author{Linda E. Reichl\footnotemark[1] \and Minh-Binh Tran\footnotemark[2] 
}

\renewcommand{\thefootnote}{\fnsymbol{footnote}}

\footnotetext[1]{Department of Physics and  Center for Complex Quantum Systems, University of Texas-Austin, Austin, TX 78712, USA. Email: reichl@mail.utexas.edu. 
}

\footnotetext[2]{Department of Mathematics, University of Wisconsin-Madison, Madison, WI 53706, USA. Email: mtran23@wisc.edu
}

\maketitle
\begin{abstract} We review recent work on a kinetic model for very low temperature dilute Bose gases. The brief derivation, expressions for hydrodynamics modes, and the comparison  with a  experiment on a BEC of $^{87}{\rm Rb}$ atoms are presented.
\end{abstract}



\tableofcontents
\section{Introduction}

After the first observation of Bose-Einstein condensation in a gas of rubidium $^{87}{\rm Rb}$ atoms  \cite{WiemanCornell} and then in a gas of sodium $^{23}{\rm Na}$ atoms  \cite{Ketterle}, there have been a number of experiments investigating properties of dilute Bose-Einstein condensates (BECs), and it has been shown that the  mean field theory of dilute BECs, first proposed by Bogoliubov \cite{Bogoliubov:SIS:1962},  gives excellent agreement with experiments \cite{dodd1998collective,hutchinson1997finite, morgan2000gapless}.

If we consider a ideal gas of bosons  at temperature $T$, as the temperature of the gas is lowered, there is a critical temperature $T_c$ at which a phase transition occurs (Bose-Einstein condensation) and bosons begin to condense into the ground  state (lowest energy state) of the system. This phase transition also occurs for weakly interacting particles like  $^{87}{\rm Rb}$ and  $^{23}{\rm Na}$, which are bosons. For finite temperatures $T<T_c$, the gas can be viewed as a two-fluid system, consisting of an irrotational fluid (the condensate) and a normal fluid (the excited particles) \cite{dalfovo1999theory,pethick2002bose,pitaevskii2016bose,Spohn:2010:KOT}. The original experiments on BECs confined the atoms to a harmonic optical trap formed  by electromagnetic fields. More recently, BECs have also been formed on microelectronic chips \cite{hansel2001bose}.

At finite temperatures, the coupling of the condensate and noncondensate degrees of freedom  leads to a two-component condensate-thermal cloud system.  The dynamical description of such systems at finite temperature  involves a kinetic equation for the dynamics of thermal excitations coupled to the evolution equation of the macroscopic phase of the Bose-Einstein condensate.   The thermal excitations described by Bogoliubov mean field theory can be regarded as a gas of weakly-interacting  excitations (``bogolons"), whose  energy spectrum is phonon-like at low temperature and particle-like at higher temperatures.

In the pioneering work by Kirkpatrick and Dorfman \cite{KD1,KD3,KD2} and Eckern \cite{E}, the authors derived a closed kinetic equation for the quasiparticle distribution function of an inhomogeneous Bose gas below the transition temperature. The approach was then employed and extended by Zaremba, Nikuni and Griffin \cite{ZarembaNikuniGriffin:1999:DOT}.  These theories can be regarded as the consistent time-dependent extension of the Hartree-Fock-Bogoliubov-Popov theory  \cite{GriffinNikuniZaremba:2009:BCG} in which  collisions within the thermal cloud and particle-exchange collisions between condensate and thermal atoms are included. Based on  a quantum BBGKY hierarchy argument, a similar model was also derived in the work \cite{PomeauBrachetMetensRica}.

Independently, using  a field-theoretic formulation of the
non-equilibrium Keldysh theory \cite{danielewicz1984quantum}, within the many-body T-matrix approximation, Stoof also derived  a model \cite{Stoof:1999:CVI} that describes the evolution of the full probability distribution
for a weakly interacting Bose gas. In this model, using a Hartree-Fock-like ansatz, the total probability distribution can be  separated into a product of respective probability distributions for the condensate
and thermal particles, that finally leads  to a similar system  to the one obtained in \cite{ZarembaNikuniGriffin:1999:DOT}. 

Based on techniques established in the quantum optics community \cite{gardiner2004quantum},   Gardiner, Zoller and co-workers \cite{QK1,QK2,QK3,QK4,QK5} developed a theory that gives a unified description of the entire range of Bose gas kinetics, combining both coherent and incoherent processes. Using the above theory, the authors also wrote a different series of papers 
\cite{gardiner1997kinetics,gardiner1998role,gardiner1998quantum,lee2000quantum,QK7},
in which the {\it formation} of a BEC 
in an optical trap was studied. The  theories formulated by Stoof \cite{duine2001stochastic,stoof1997initial,Stoof:1999:CVI},  and those formulated by Gardiner-Zoller \cite{QK1,QK3,QK5,QK4,QK2}, both have the kinetic equations of ZNG as a limiting case. 
An excellent review of some of these kinetic theories  can be found in  \cite{proukakis2013quantum}.

All of the above models are based on a picture of excitations with particle-like spectrum. Such models are  adequate for high temperature ranges  $T_{BEC}>T\ge 0.5T_{BEC}$ \cite{IG}, but are inappropriate for describing collective phonon-like excitations, which become important at very low temperatures \cite{allen2012dynamical}. In order to fix this problem, in \cite{ReichlGust:2013:TTF,ReichlGust:2013:RRA,ReichlGust:2012:CII,gust2012relaxation,ReichlGust:2014:DOH,gust2015viscosity}  authors Reichl and Gust, based on the work of Peletminskii and Yatsenko \cite{PeletminskiiYatsenko:1968:FTB}, derived a new kinetic equation which takes into account the non-conservation of bogolon number during collisions and the phonon-like spectrum of bogolons at  very low temperature. As a result,   a new contribution to  the collision operator $\mathcal{G}^{31}$ appears that takes into account $1{\leftrightarrow}3$ type collisions between the excitations, addition to the $1{\leftrightarrow}2$ and $2{\leftrightarrow}2$ type collisions that are known to occur.  In \cite{PeletminskiiYatsenko:1968:FTB}, Peletminskii and Yatsenko derived a more traditional kinetic equation that could incorporate a mean field description of relaxation processes insuperfluids. This approach was subsequently used to describe relaxation processes in Fermi superfluids \cite{Reichl:MMI1:1980, Reichl:MMI2:1980} and later used to derive  the kinetic equations BECs that are discussed in more detail in subsequent sections.

In this review paper, we revisit the model derived by Reichl and Gust \cite{ReichlGust:2013:TTF,ReichlGust:2013:RRA,ReichlGust:2012:CII,gust2012relaxation,ReichlGust:2014:DOH,gust2015viscosity}. The model, which couples the  kinetic equation for bogolons to the equation for evolution of the condensate, is described in Section 2. In Section 3, we recall the main steps of the derivation of the model by Reichl and Gust. In Section 4, we  calculate the decay rates of the sound modes as a function equilibrium temperature, density, particle mass and interaction strength. We obtain expressions for the decay rates of sound modes that can be applied to any monatomic dilute Bose gas and compare the results to a  experiment on a BEC of $^{87}{\rm Rb}$ atoms \cite{steinhauer12}. The value of the sound mode lifetime, predicted by the new theory, is consistent with experiment reported in \cite{steinhauer12}.

\section{The model}

In this paper, we are interested in the kinetic equations that describe  the dynamics of excitations (bogolons) in a very low temperature dilute atomic Bose-Einstein condensate  \cite{ReichlGust:2013:TTF,ReichlGust:2013:RRA,ReichlGust:2012:CII,gust2012relaxation,ReichlGust:2014:DOH,gust2015viscosity}.  We let $f({\bf r},\kk_1,t)$ denote the deviation from equilibrium of the phase space bogolon number density for bogolons,  at time $t$,  with position and momentum in the intervals ${\bf r}{\rightarrow}{\bf r}+d{\bf r}$, ${\hbar}{\bf k}_1{\rightarrow}{\hbar}{\bf k}_1+d{\hbar}{\bf k}_1$, respectively (${\hbar}$ is Planck's constant).  Then the equation for the spatial Fourier transform, 
$f({\bf q},\kk_1,t)={\int}d{\bf r}{\rm e}^{i{\bf q}{\cdot}{\bf r}} f({\bf r},\kk_1,t)$, satisfies the coupled equations
\begin{eqnarray}
\nonumber
\frac{{\partial}f( {\bf q},\kk_1,t)}{{\partial}t} & =& i\frac{{\hbar}}{m}\kk_1{\cdot} {\bf q} \frac{{\epsilon}(\kk_1)+{\Lambda_0}}{E_{\kk_1}}f({\bf q},\kk_1,t)
+i {\bf q} {\cdot}{\bf v}_s({\bf q},t){\mathcal N}^{\rm eq}_{\kk_1}-\\\label{kineq1a}
&& \ \ \ \ \ \    - {\bf G}{[}f{]({\bf q},\kk_1,t)}, \mbox{ on }  ({\bf q},\kk_1,t)\in \bf{R}^3\times\bf{R}^3\times \bf{R}_+,\\\label{kineq1b}
f({\bf q},\kk_1,0) & =& f_0({\bf q},\kk_1),\ \ \ \ ({\bf q},\kk_1)\in \bf{R}^3\times\bf{R}^3,\\\nonumber
\frac{{\partial}^2{\phi}({{\bf q}},t)}{{\partial}t^2}
& = &-i\frac{g}{m}~\frac{1}{(2{\pi})^3}{\int}_{\bf{R}^3}d{\bf k}_1~{{\bf q}}{\cdot}{\bf k}_1f({\bf q},\kk_1,t) -\\\label{phi1a}
& &  \ \ \ \ \   -i\frac{g}{\hbar}{{\bf q}}{\cdot}{\bf v}_s({{\bf q}},t){n}^{\rm eq}, \mbox{ on }  ({\bf q},t) \in \bf{R}^3\times \bf{R}_+,\\\label{phi1b}
\phi({\bf q},0) & =& \phi_0({\bf q}),\ \ \ \ \xi \in \bf{R}^3,
\end{eqnarray}
where  ${\bf q}$ is the wave vector for spatial variations of the bogolon density, ${\epsilon}_{\kk_1}=\frac{{\hbar}^2|\kk_1|^2}{2m}$, $\Lambda_0$ is the equilibrium condensate  order parameter,
$$E_{\kk_1}=\sqrt{({\epsilon}_{\kk_1}+{\Lambda_0})^2-{\Lambda_0}^2}$$ 
is the bogolon energy, and 
$${\mathcal N}^{\rm eq}_{{\bf k}_1}=({\rm e}^{E_{\kk_1}/k_BT}-1)^{-1}$$
 is the equilibrium Bose-Einstein distribution for  bogolons at temperature $T$ with $k_B$ being the Boltzmann constant, $g=4{\pi}{\hbar}^2a/m$ is the coupling constant, $a$ is the s-wave scattering length of the atoms in the gas, and ${n}^{\rm eq}$ is the total particle number density.
 The distribution  ${\mathcal N}^{\rm eq}({\bf k}_1)$ is a stationary state of  Eq. (\ref{kineq1a}).

The macroscopic phase of the condensate, ${\phi}({\bf r},t)$, varies in space and time. The equation for the  component  ${\phi}({\bf q},t)={\int}d{\bf r}{\rm e}^{i{\bf q}{\cdot}{\bf r}} {\phi}({\bf r},t)$ with wave vector ${\bf q}$
is given by Eq. (\ref{phi1a}).  The equilibrium particle density $n^{\rm eq}$ that appears in Eq. (\ref{phi1a}) can be written
\begin{equation}
{n}^{\rm eq}{\approx}~{n}_0^{\rm eq}+\frac{1}{(2{\pi})^3}{\int}_{\bf{R}^3}d{\bf k}_1\frac{{\epsilon}_{\kk_1}+\Lambda_0}{E_{\kk_1}}{\N}^{\rm eq}_{\kk_1}
\label{AA}
\end{equation}
where ${\rm n}_0^{eq}$ is the density of particles that have  condensed into the ground state  $k=0$. This form of the equilibrium particle density is sometimes called the "Popov approximation" and  limits the theory to temperatures below about $0.6T_C$ \cite{hutchinson1997finite}.   The  superfluid velocity is determined by the spatial variation of the macroscopic phase ${\phi}({\bf r},t)$ and is given by  by ${\bf v}_s({\bf r},t)=\frac{{\hbar}}{m}{\nabla}_{\bf r}{\phi}({\bf r},t)$. Therefore,
\begin{equation}
{\bf v}_s({\bf q},t)=-i\frac{\hbar}{m}{{\bf q}}~{\phi}({{\bf q}},t).\end{equation}
The macroscopic phase  ${\phi}({\bf r},t)$ arises from the broken gauge symmetry in the Bose Einstein condensate. From Eqs. (\ref{kineq1a}). -(\ref{phi1b}), we see that bogolon distribution function and the macroscopic phase are nonlinearly coupled.

The collision operator $\bf{G}{[}f{]}$ that appears in Eq. (\ref{kineq1a}) contains the processes that cause the BEC to relax to equilibrium.  In subsequent sections, we consider the linearized bogolon kinetic equation so we write the linearized collision operator here.   Let us define
\begin{equation}
f({\bf q},\kk,t)={\N}^{\rm eq}_{\kk}(1+{\N}^{\rm eq}_{\kk})\eta({\bf q},\kk,t),
\label{small}
\end{equation}
then $\eta({\bf q},{\bf k},t)$ is a small quantity that decays to zero as the gas relaxes to equilibrium. We rewrite $\bf{G}{[}f{]}$ as ${\mathcal G}{[}\eta {]}$.  In the hydrodynamic regime where spatial variations have very long wavelength, the wave vector 
$|{\bf q}|$ is a very small parameter.
The linearized bogolon collision integral can be written  ${\mathcal G}{[}\eta {]}={\mathcal G}^{12}{[}\eta {]}+{\mathcal G}^{22}{[}\eta {]}+{\mathcal G}^{13}{[}\eta {]}$ where 
\begin{eqnarray}
\mathcal{G}^{12}_{{\bf  k}_1,{\bf q}} = \frac{4 \pi N_0 g^2}{\hbar V^2} {\sum_{2,3}}' {\delta}^{(4)}_{1,2+3}(W^{12}_{3,2,1} )^2 {\mathcal M}^{eq}_1 \N^{eq}_2 \N^{eq}_3 ({\eta}_2+{\eta}_3-{\eta}_1)~~~\nonumber\\
+ \frac{8 \pi N_0 g^2}{\hbar V^2} {\sum_{2,3}}' \delta^{(4)}_{1+2,3} (W^{12}_{1,2,3})^2~
{\mathcal M}^{eq}_3 \N^{eq}_1 \N^{eq}_1 ({\eta}_3-{\eta}_1-{\eta}_2),~~~
\end{eqnarray}
\begin{eqnarray}
\mathcal{G}^{22}_{{\bf  k}_1,{\bf q}} = \frac{4 \pi g^2}{\hbar V^2} {\sum_{2,3,4}}'
\delta^{(4)}_{1+2,3+4} ( W^{22}_{1,2,3,4})^2 ~{\mathcal M}^{eq}_1{\mathcal M}^{eq}_2 \N^{eq}_3 \N^{eq}_4 ~~~~~~~~~~~~~~\nonumber\\
{\times}({\eta}_3+{\eta}_4-{\eta}_1-{\eta}_2)~~~~~~~~~
\end{eqnarray}
and
\begin{eqnarray}
{\delta}\mathcal{G}^{31}_{{\bf  k}_1,{\bf q}} = \frac{4 \pi g^2}{3 \hbar V^2} {\sum_{2,3,4}}' \delta^{(4)}_{1,2+3+4} (W^{31}_{1,2,3,4} )^2{\mathcal M}^{eq}_1\N^{eq}_2 \N^{eq}_3 \N^{eq}_4~~~~~~~~~~\nonumber\\
{\times} ({\eta}_2+{\eta}_3+{\eta}_4-{\eta}_1)~~~~~~~~~~\nonumber\\
+\frac{4 \pi g^2}{\hbar V^2} {\sum_{2,3,4}}' \delta^{(4)}_{1+2+3,4} (W^{31}_{4,3,2,1} )^2
 \N^{eq}_1\N^{eq}_2 \N^{eq}_3{\mathcal M}^{eq}_4 ~~~~~~~~~~~~\nonumber\\
 {\times} ({\eta}_4-{\eta}_1-{\eta}_2-{\eta}_3)~~~~~~~~~~
\end{eqnarray}
where ${\mathcal M}^{eq}_j=1+\N^{eq}_j$,   ${\eta}_j={\eta}({\bf q},{\bf k}_j,t)$, $E_i=E_{{\bf k}_i}$, the summation ${\sum}_j^{'}={\sum_{{\bf k}_j{\neq}0}}$,  and ${\delta}^{(4)}_{1+2,3+4}$ denotes the product of momentum and energy conserving delta functions
\begin{eqnarray}
{\delta}^{(4)}_{1+2,3+4}={\delta}^{(3)}({\bf k}_1+{\bf k}_2-{\bf k}_3-{\bf k}_4){\delta}(E_1+E_2-E_3-E_4).
\end{eqnarray}

The weighting functions $W^{12}_{1,2,3} $, $W^{22}_{1,2,3,4}$, and $W^{31}_{1,2,3,4} $,   are given by
\begin{equation}
W^{12}_{1,2,3} = u_1 u_2 u_3 - u_1 v_2 u_3 - v_1 u_2 u_3 + u_1 v_2 v_3 + v_1 u_2 v_3 - v_1 v_2 v_3,
\end{equation}
\begin{equation}
W^{22}_{1,2,3,4} = u_1 u_2 u_3 u_4 + u_1 v_2 u_3 v_4 + u_1 v_2 v_3 u_4 + v_1 u_2 u_3 v_4 + v_1 u_2 v_3 u_4 + v_1 v_2 v_3 v_4
\end{equation}
and
\begin{equation}
W^{31}_{1,2,3,4} = u_1 u_2 u_3 v_4 + u_1 u_2 v_3 u_4 + u_1 v_2 u_3 u_4 + v_1 v_2 v_3 u_4 + v_1 v_2 u_3 v_4 + v_1 u_2 v_3 v_4.
\end{equation}
 In the limit that $\Lambda_0 \to 0$, the weighting functions $W^{12}$ and $W^{22}$ approach $1$ while the weighting function $W^{31}$ approaches zero. However, $\mathcal{G}^{12}$ still approaches zero overall since it is multiplied by $N_0$ which approaches zero. The collision operators $\mathcal{G}^{12}$ and ${\cal G}^{22}$ are the same as those considered in refs. \cite{KD1,KD2,KD3,GriffinNikuniZaremba:2009:BCG} . They dominate collision processes at higher temperatures where particle-like excitations exist. The collision operator $\mathcal{G}^{31}$ becomes important at lower temperatures where the excitations take on a more  phonon-like character.The  factors $u_i$ and $v_i$ are given by
\begin{equation}
u_i = \frac{1}{\sqrt{2}}\sqrt{\frac{\epsilon_i +{\Lambda}_0}{E_i} + 1} \hspace{1in} v_i = \frac{1}{\sqrt{2}}\sqrt{\frac{\epsilon_i + {\Lambda}_0}{E_i} - 1}
\end{equation}
where ${\epsilon}_j={\epsilon}_{k_j}$. In Appendix A, we write these linearized collision operators in a more explicit form.

\section{Derivation of the system}

In this Section, we outline the key steps of the derivation (cf. \cite{ReichlGust:2013:TTF,ReichlGust:2012:CII}) of the BEC kinetic equations shown in Section 2. Let ${\hat \Phi}^{\dagger}({\bf x})$ ($ {\hat \Phi}({\bf x})$) be the quantum field operator that creates (annihilates) a particle at position ${\bf x}$, 
the Hamiltonian for  $N$ bosons of mass $m$, in a  cubic box   with very large volume $\Omega$,  can be written in the form 
\begin{eqnarray}
{\hat H}=\int_{\Omega} d{\bf x} {\hat \Phi}^{\dagger}({\bf x}) \left(-\frac{{\hbar}^2}{2m}\Delta_{\bf x}\right){\hat \Phi}({\bf x})~~~~~~~~~~~~~~~~~~~~~~\nonumber\\
+\frac{1}{2}{\int}_{\Omega}{\int}_{\Omega}d{\bf x}_1d{\bf x}_2 {\rm V}(|{\bf x}_1-{\bf x}_2|) {\hat \Phi}^{\dagger}({\bf x}_1) {\hat \Phi}^{\dagger}({\bf x}_2) {\hat \Phi}({\bf x}_2) {\hat \Phi}({\bf x}_1),
\label{hamexact1}
\end{eqnarray}
where   and the integration is over the entire volume $\Omega$ of the cubic box.  These operators satisfy
the boson commutation relations
$[ {\hat \Phi}({\bf x}_1),{\hat \Phi}^{\dagger}({\bf x}_2)]={\delta}({\bf x}_1-{\bf x}_2)$. We will assume that the interaction between particles is given by the contact potential
$ {\rm V}(|{\bf x}_1-{\bf x}_2|)=g{\delta}({\bf x}_1-{\bf x}_2)$. 
The evolution of the probability density operator ${\hat \rho}$ for this system follows the quantum Liouville equation
\begin{equation}
{{\partial}{\hat \rho}(t)\over {\partial}t}=-{i\over \hbar}[{\hat H},{\hat \rho}(t)].
\label{liouexact1}
\end{equation}
The equations (\ref{hamexact1}) and (\ref{liouexact1}) give the exact behavior of the BEC gas.

Below the Bose-Einstein condensation transition temperature, $T_c$, the gauge symmetry of the fluid is broken. In order to accurately describe the behavior of the BEC, one needs to   incorporate this broken symmetry into the dynamics, by using the one-body reduced density operator 
\begin{equation}
{\hat {\bar \Theta}}({\bf x}_1,{\bf x}_2)=\left(
\begin{array}{cc}
{\hat  \Phi}^{\dagger}({\bf x}_1) {\hat \Phi}({\bf x}_2) &  {\hat \Phi}^{\dagger}({\bf x}_1) {\hat \Phi}^{\dagger}({\bf x}_2)\\
{\hat   \Phi}({\bf x}_1) {\hat \Phi}({\bf x}_2) &  {\hat \Phi}({\bf x}_1) {\hat \Phi}^{\dagger}({\bf x}_2)\\
\end{array}\right),
\end{equation}
and the  one-body reduced density matrix  %
\begin{eqnarray}
{\bar {\mathcal F}}({\bf x}_1,{\bf x}_2,t)=
{\rm Tr}\left[{\hat \rho}(t){\hat {\bar \Theta}}({\bf x}_1,{\bf x}_2)\right]
=\left( \begin{array}{cc}
{\langle}{\hat  \Phi}^{\dagger}({\bf x}_1) {\hat \Phi}({\bf x}_2){\rangle} & {\langle} {\hat \Phi}^{\dagger}({\bf x}_1) {\hat \Phi}^{\dagger}({\bf x}_2){\rangle}\\
{\langle}{\hat   \Phi}({\bf x}_1) {\hat \Phi}({\bf x}_2){\rangle} &{\langle}  {\hat \Phi}({\bf x}_1) {\hat \Phi}^{\dagger}({\bf x}_2){\rangle}\\
\end{array}\right),~~~~~~~~~~~~~~~~~~~~
\end{eqnarray}
which follows the time evolution equation
\begin{equation}
-i{\hbar}\frac{{\partial}{{{\bar {\mathcal F}}}({\bf x}_1,{\bf x}_2,t)}}{{\partial}t}={\rm Tr}[{\hat \rho}(t)~[{\hat H},{\hat {\bar \Theta}}({\bf x}_1,{\bf x}_2)]]
\end{equation}

According to the Bogoliubov assumption, after a very short time $t$    the density operator 
${\hat \rho}(t)$ will be a functional of the single particle reduced density operator
${\bar {\mathcal F}}({\bf x}_1,{\bf x}_2,t)$. The density operator then 
can be written, with an abuse of notation
\begin{equation}
{\hat \rho}(t)={\hat { \rho}'}({\{}{\bar {\mathcal F}}{\}}),
\end{equation}
where ${\{}{\bar {\mathcal F}}{\}}$ denotes the vector containing ${\bar {\mathcal F}}({\bf x}_1,{\bf x}_2,t)$ for all values of $({\bf x}_1,{\bf x}_2)$.
The quantity ${\bar {\mathcal F}}({\bf x}_1,{\bf x}_2,t)$ is defined self-consistently so that
\begin{equation}
{\bar {\mathcal F}}({\bf x}_1,{\bf x}_2,t)={\rm Tr}[{\hat \rho}'({\{}{\bar {\mathcal F}}{\}}) {\hat {\bar  \Theta}}({\bf x}_1,{\bf x}_2)].
\label{density}
\end{equation}

The existence of the broken symmetry can be made explicit   if we divide the total Hamiltonian  into a mean field contribution  ${\hat H}_0$ and a deviation from the mean field ${\hat H}_1$. The total Hamiltonian then  takes the form ${\hat H}={\hat H}_0+{\hat H}_1$, where the mean field Hamiltonian  is defined
\begin{equation}
{\hat H}_0=\int_{\Omega}d{\bf x}{\hat  \Phi}^{\dagger}({\bf x}) \left(-\frac{{\hbar}^2}{2m}\Delta_{\bf x} -{\mu}\right) {\hat \Phi}({\bf x})+{\hat H}_3,
\label{ham0mf}
\end{equation}
with
\begin{eqnarray}
{\hat H}_3=\frac{1}{2}{\int_B}d{\bf x}_1 [{\mathcal \nu}({\bf x}_1) {\hat \Phi}^{\dagger}({\bf x}_1) {\hat  \Phi}({\bf x}_1)+{\mathcal \nu}({\bf x}_1){\hat  \Phi}({\bf x}_1) {\hat  \Phi}^{\dagger}({\bf x}_1)] \nonumber\\
+\frac{1}{2}{\int_{\Omega}}d{\bf x}_1{\mathbf \Lambda}^{\dagger}({\bf x}_1){\hat  \Phi}({\bf x}_1) {\hat \Phi}({\bf x}_1)
+\frac{1}{2}{\int_{\Omega}}d{\bf x}_1{\mathbf \Lambda}({\bf x}_1){\hat  \Phi}^{\dagger}({\bf x}_1){\hat  \Phi}^{\dagger}({\bf x}_1),
\label{hamUmf}
\end{eqnarray}
and ${ \hat H}_1$ contains deviations   from the mean field Hamitloniain
\begin{equation}
{ \hat H}_1=\frac{1}{2}{\int_{\Omega}}{\int_{\Omega}}d{\bf x}_1d{\bf x}_2 {\rm V}(|{\bf x}_1-{\bf x}_2|) {\hat \Phi}^{\dagger}({\bf x}_1)  {\hat \Phi}^{\dagger}({\bf x}_2) {\hat \Phi}({\bf x}_2){\hat  \Phi}({\bf x}_1) - {\hat H}_3,
\label{ham1mf}
\end{equation}
In Eq. (\ref{hamUmf}),  ${\mathcal \nu}({\bf x}_1)=2g~{\langle}{\hat  \Phi}^{\dagger}({\bf x}_1) {\hat \Phi}({\bf x}_1){\rangle}$,  
${\Lambda}({\bf x}_1)=g~{\langle}{\hat  \Phi}({\bf x}_1) {\hat \Phi}({\bf x}_1){\rangle}$, 
${ \Lambda}^{\dagger}({\bf x}_1)=g~{\langle}{\hat  \Phi}^{\dagger}({\bf x}_1) {\hat \Phi}^{\dagger}({\bf x}_1){\rangle}$, and ${\mu}$ is the equilibrium chemical potential.
\subsection{The Kinetic Equation}

In \cite{ReichlGust:2012:CII}, the authors  used the Peletminksii and Yatsenko approach 
\cite{AkhiezerPeletminski:SOS:1981,PeletminskiiYatsenko:1968:FTB} to derive the  kinetic equation for the BEC from the above
mean field Hamiltonian. The kinetic equation describing the dynamic evolution of the one-body density matrix can be written

\begin{eqnarray}
-i{\hbar}\frac{{\partial}{{\bar {\mathcal F}}({\bf x}_1,{\bf x}_2,t)}}{{\partial}t}={\rm Tr}{\{}{\hat \rho}'({\{}{\bar {\mathcal F}}{\}}),[{\hat H}_0,{\hat { \bar  \Theta}}({\bf x}_1,{\bf x}_2)] {\}}+{\rm Tr}{\{}{\hat \rho}'({\{}{\bar {\mathcal F}}{\}}),[{\hat  H}_1,{\hat {\bar \Theta}}({\bf x}_1,{\bf x}_2)]{\}}   \nonumber\\
+\frac{i}{\hbar}{\int_{-\infty}^0}ds~{\rm Tr}{\{}{\hat \rho}'({\{}{\bar {\mathcal F}}{\}}),[{\hat  H}_1,{\hat S}^{0,\dagger}(0,s)[{\hat  {\bar \Theta}}({\bf x}_1,{\bf x}_2),{\hat  H}_1]{\hat S}^{0}(0,s){\}},
\end{eqnarray}
where ${\hat S}^{0}$ is the semigroup operator
\begin{equation}
{\hat S}^{0}(s_1,s_2)={\rm e}^{-{\hat H}_0(s_1-s_2)/{\hbar}},
\end{equation}
and ${\hat S}^{0,\dagger}$ is the adjoint of ${\hat S}^{0}$.
The mean  field Hamiltonian ${\hat H}_0$, defined in (\ref{ham0mf})  needs to satisfy 
\begin{equation}
{\rm Tr}{\{}{\hat \rho}'({\{}{\bar {\mathcal F}}{\}})[{\hat  H}_1,{\hat { \Theta}}({\bf x}_1,{\bf x}_2)]{\}} =0,
\label{density}
\end{equation}
In order to remove secular effects in the evolution of the one-body density matrix.

We can  now introduce the unitary transformation to the reference frame moving with the superfluid (superfluid rest frame)
\begin{equation}
{\hat S}(t)={\rm exp}\left[ -i{\int_{\Omega}}d{\bf x}{\phi}({\bf x},t){\hat  \Phi}^{\dagger}({\bf x}){\hat  \Phi}({\bf x})\right],
\end{equation}
where ${\phi}({\bf x},t)$ is the macroscopic phase of the condensate wave function. 
We let ${\hat  \psi}^{\dagger}({\bf x}) $ and ${\hat  \psi}({\bf x}) $ denote particle creation and annihilation operators in the superfluid rest frame. Then
${\hat S}^{\dagger}(t){\hat  \Phi}({\bf x}){\hat S}(t)={\rm e}^{-i{\phi}({\bf x},t)}{\hat  \Phi}({\bf x})={\hat  \psi}({\bf x})$  
and we obtain
\begin{equation}\begin{aligned}
-i{\hbar}\frac{{\partial}}{{\partial}t} {\langle}{\hat  \psi}^{\dagger}_1 {\hat \psi}_2{\rangle}&=(\mathcal{L}_1^{(+)}-\mathcal{L}_2^{(-)}){\langle}{\hat  \psi}^{\dagger}_1 {\hat \psi}_2{\rangle}-{\Lambda}_2{\langle}{\hat  \psi}^{\dagger}_1 {\hat \psi}^{\dagger}_2{\rangle}+{\Lambda}^{\dagger}_1{\langle}{\hat  \psi}_1 {\hat \psi}_2{\rangle}+{\mathcal I}_{11}, \\
-i{\hbar}\frac{{\partial}}{{\partial}t} {\langle}{\hat  \psi}^{\dagger}_1 {\hat \psi}^{\dagger}_2{\rangle}&=(\mathcal{L}_2^{(+)}+\mathcal{L}_1^{(+)}){\langle}{\hat  \psi}^{\dagger}_1 {\hat \psi}^{\dagger}_2{\rangle}+{\Lambda}^{\dagger}_2{\langle}{\hat  \psi}^{\dagger}_1 {\hat \psi}_2{\rangle}+{\Lambda}^{\dagger}_1{\langle}{\hat  \psi}_1 {\hat \psi}^{\dagger}_2{\rangle} +{\mathcal I}_{12}, \\
-i{\hbar}\frac{{\partial}}{{\partial}t} {\langle}{\hat  \psi}_1 {\hat \psi}_2{\rangle}&=-(\mathcal{L}_2^{(-)}+\mathcal{L}_1^{(-)}){\langle}{\hat  \psi}_1 {\hat \psi}_2{\rangle}-{\Lambda}_1{\langle}{\hat  \psi}^{\dagger}_1 {\hat \psi}_2{\rangle}-{\Lambda}_2{\langle}{\hat  \psi}_1 {\hat \psi}^{\dagger}_2{\rangle}+{\mathcal I}_{21},  \\
-i{\hbar}\frac{{\partial}}{{\partial}t} {\langle}{\hat  \psi}_1 {\hat \psi}^{\dagger}_2{\rangle}&=(\mathcal{L}_2^{(+)}-\mathcal{L}_1^{(-)}){\langle}{\hat  \psi}_1 {\hat \psi}^{\dagger}_2{\rangle}-{\Lambda}_1{\langle}{\hat  \psi}^{\dagger}_1 {\hat \psi}^{\dagger}_2{\rangle}+{\Lambda}^{\dagger}_2{\langle}{\hat  \psi}_1 {\hat \psi}_2{\rangle}+{\mathcal I}_{22},  
\label{kineq1c}\end{aligned}
\end{equation}
where $\hat\psi_j  = \hat\psi({\bf x}_j)$, 
\begin{equation}
\mathcal{L}^{(\pm)}_j={L}({\bf x}_j) {\pm}i\frac{{\hbar}}{2}({\nabla}_{{\bf x}_j}{\cdot}{\bf v}_s({\bf x}_j))+\frac{m}{2}{\bf v}^2_s({\bf x}_j)~ {\pm}~i{\hbar}{\bf v}_s({\bf x}_j){\cdot}{\nabla}_{{\bf x}_j}+{\hbar}\frac{{\partial}{\phi}({{\bf x}_j})}{{\partial}t},
\end{equation}
with
\begin{equation}\begin{aligned}
{L}({\bf x}_j)&=-\frac{{\hbar}^2}{2m}\Delta_{{\bf x}_j}+{\nu}({\bf x}_j)-{\mu},~~~~
{\nu}({\bf x}_j)=2g{\langle}{\hat \psi}^{\dagger}({\bf x}_j){\hat \psi}({\bf x}_j){\rangle},\\
~{\Lambda}_j&={\Lambda}({\bf x}_j)=g{\langle}{\hat \psi}({\bf x}_j){\hat \psi}({\bf x}_j){\rangle}, ~~~~{\Lambda}^{\dagger}_j ={\Lambda}^{\dagger}({\bf x}_j)=g{\langle}{\hat \psi}^{\dagger}({\bf x}_j){\hat \psi}^{\dagger}({\bf x}_j){\rangle},\end{aligned}
\end{equation}
for $j=1,2$.
The quantity $~{\bf v}_s({\bf x}_j)=\frac{{\hbar}}{m}{\nabla}_{{\bf x}_j}{\phi}({\bf x}_j)$ is the superfluid velocity. 
The quantities 
\begin{equation}
\left( \begin{array}{cc}
 {\mathcal I}_{1,1}&  {\mathcal I}_{1,2}\\
{\mathcal I}_{2,1} &  {\mathcal I}_{2,2}\\
\end{array}\right)=\frac{i}{\hbar}{\int_{-\infty}^0}ds~{\rm Tr}{\{}{\hat \rho}'({\{}{\bar {\mathcal F}}{\}})[{\hat  H}_1,{\hat S}^{0,\dagger}(0,s)[{\hat {\bar \theta}}({\bf x}_1,{\bf x}_2),{\hat  H}_1]{\hat S}^{0}(0,s)]{\}},
\end{equation}
where 
\begin{equation}
{\hat {\bar \theta}}({\bf x}_1,{\bf x}_2)=\left(
\begin{array}{cc}
{\hat  \psi}^{\dagger}({\bf x}_1) {\hat \psi}({\bf x}_2) &  {\hat \psi}^{\dagger}({\bf x}_1) {\hat \psi}^{\dagger}({\bf x}_2)\\
{\hat   \psi}({\bf x}_1) {\hat \psi}({\bf x}_2) &  {\hat \psi}({\bf x}_1) {\hat \psi}^{\dagger}({\bf x}_2)\\
\end{array}\right),
\end{equation}
are the collision integrals  governing  relaxation processes in the BEC gas. 

The coupled kinetic equations (\ref{kineq1c}) contain the full quantum dynamics of the BEC gas.  If we transform these kinetic equations  to equations for the Wigner functions,  we can write the kinetic equations in  the hydrodynamic regime where all macroscopic quantities are slowly varying in space and time.

\subsection{Kinetic Equations in Terms of Wigner Functions}

Wigner functions are distribution functions  in phase space for  quantum  systems \cite{wigner}. They are particularly useful in dealing with transport processes because in the classical limit they reduce to classical probability distributions in phase space. The field operators ${\hat  \psi}^{\dagger}_1$ and $ {\hat \psi}_1$  are related to operators ${\hat  a}^{\dagger}_{{\bf k}_1}$ and ${\hat a}_{{\bf k}_1}$,  that create and annihilate, respectively, a particle with momentum ${\hbar}{\bf k}_1$,  via the Fourier transforms
\begin{equation}
{\hat  \psi}^{\dagger}_1=\frac{1}{\sqrt{\Omega}}{\sum_{{\bf k}_1}} {\rm e}^{-i{\bf k}_1{\cdot}{\bf r}_1}{\hat  a}^{\dagger}_{{\bf k}_1}, ~~{\rm and }~~{\hat  \psi}_1=\frac{1}{\sqrt{\Omega}}{\sum_{{\bf k}_1}} {\rm e}^{+i{\bf k}_1{\cdot}{\bf r}_1}{\hat  a}_{{\bf k}_1}. 
\end{equation}
We can therefore relate the configuration space distributions to momentum space distributions via the Fourier transformation
\begin{equation}
\left( \begin{array}{cc}
{\langle}{\hat  \psi}^{\dagger}_1 {\hat \psi}_2{\rangle} & {\langle} {\hat \psi}^{\dagger}_1 {\hat \psi}^{\dagger}_2{\rangle} \\
{\langle}{\hat   \psi}_1 {\hat \psi}_2{\rangle}&
{\langle}  {\hat \psi}_1 {\hat \psi}^{\dagger}_2{\rangle} \\
\end{array}\right)=\frac{1}{\Omega}{\sum_{{{\bf k}_1},{{\bf k}_2}}} {\rm e}^{-i{\bf k}_1{\cdot}{\bf r}_1}{\rm e}^{+i{\bf k}_2{\cdot}{\bf r}_2}\left( \begin{array}{cc}
{\langle}{\hat  a}^{\dagger}_{{\bf k}_1} {\hat a}_{{\bf k}_2}{\rangle} & {\langle}{\hat  a}^{\dagger}_{{\bf k}_1} {\hat a}^{\dagger}_{-{\bf k}_2}{\rangle} \\
{\langle}{\hat  a}_{-{\bf k}_1} {\hat a}_{{\bf k}_2} {\rangle}&
{\langle} {\hat  a}_{-{\bf k}_1} {\hat a}^{\dagger}_{-{\bf k}_2}{\rangle} \\
\end{array}\right)
\end{equation}

Let us introduce center of mass and relative coordinates ${\bf R}=\frac{1}{2}({\bf x}_1+{\bf x}_2)$ and ${\bf r}={\bf x}_1-{\bf x}_2$, respectively, 
and introduce center of mass and relative wavevectors ${\bf k}=\frac{1}{2}({\bf k}_1+{\bf k}_2)$ and ${\bf q}={\bf k}_1-{\bf k}_2$, respectively,  the Wigner functions for the BEC, whose spatial disturbance has wave vector ${\bf q}$,  are then defined 
\begin{equation}
\left( \begin{array}{cc}
F_{11}({\bf k},{\bf q}) &
 F_{12}({\bf k},{\bf q})\\
F_{21}({\bf k},{\bf q}) & 
F_{22}({\bf k},{\bf q})\\
\end{array}\right)
={\int}~d{\bf r}~{\int}~d{\bf R}~ {\rm e}^{+i{\bf k}{\cdot}{\bf r}} {\rm e}^{+i{\bf q}{\cdot}{\bf R}}  \left( \begin{array}{cc}
{\langle}{\hat  \psi}^{\dagger}_1 {\hat \psi}_2{\rangle} & {\langle} {\hat \psi}^{\dagger}_1 {\hat \psi}^{\dagger}_2{\rangle} \\
{\langle}{\hat   \psi}_1 {\hat \psi}_2{\rangle}&
{\langle}  {\hat \psi}_1 {\hat \psi}^{\dagger}_2{\rangle} \\
\end{array}\right)
\label{Wigmat5}
\end{equation}
where ${\hbar}{\bf k}$ is the momentum of particles. In the classical limit,   $F_{11}({\bf k},{\bf R}) $ is the particle number  density  in the the interval ${\bf k}{\rightarrow}{\bf k}+d{\bf k}$ and for spatial disturbances with wave vector  ${\bf q}{\rightarrow}{\bf q}+d{\bf q}$.
The particle number density whose spatial variation has wavevector ${\bf q}$ is $N({\bf q})={\sum_{\bf k}}F_{11}({\bf k},{\bf q})$. The number of particles $N({\bf k})$  with momentum ${\hbar}{\bf k}$ is
$N({\bf k})={\int}~d{\bf R}~ F_{11}({\bf k},{\bf R})={\langle}{\hat  a}^{\dagger}_{\bf k} {\hat a}_{\bf k}{\rangle}$. 
The component of the order parameters whose spatial variation has wavevector ${\bf q}$ is given by
${\Lambda}^{\dagger}({\bf q})=g{\sum_{\bf k}}F_{12}({\bf k},{\bf q})$ and ${\Lambda}({\bf q})=g{\sum_{\bf k}}F_{21}({\bf k},{\bf q})$.

 Since we are interested in the hydrodynamic regime, where all macroscopic quantities are slowly varying in space, we   keep only the lowest order derivatives with respect to ${\bf R}$ in the kinetic equations. This is equivalent to keeping only the lowest order contributions from the wave vector ${\bf q}$ (up to order $q^2$) in the kinetic equations.  

 We also note that expressions for transport coefficients can be computed from kinetic equations that are linearized about absolute equilibrium.  
 We  therefore now write the hydrodynamic variables in terms of their equilibrium values plus small perturbations from their equilibrium values, 
\begin{eqnarray}
F_{i,j}({\bf q},{\bf k})=F^{eq}_{i,j}({\bf k})+{\delta}F_{i,j}({\bf q},{\bf k}),~~~{\bf v}_s({\bf q})={\bf v}_s^0+{\delta}{\bf v}_s({\bf q})
\nonumber\\
{\Lambda}({\bf q})={\Lambda_0}+{\delta}{\Lambda}({\bf q}),~~~{\Lambda}^{\dagger}({\bf q})={\Lambda_0}+{\delta}{\Lambda}^{\dagger}({\bf q}).
\end{eqnarray}
where $F^{eq}_{i,j}({\bf k})$, ${\bf v}_s^0$, and ${\Lambda}_0$ denote the equilibrium values of the various quantities. 
We will study the  Bose gas at temperatures below $0.6T_c$, where the Popov approximation has been shown to give good agreement with experiments and $F^{eq}_{11}({\bf 0}){\approx} F^{eq}_{12}({\bf 0}) {\approx} F^{eq}_{21}({\bf 0}) {\approx} F^{eq}_{22}({\bf 0}){\approx} {\rm N}^{eq}_{\bf 0}$, with ${\rm N}^{eq}_{\bf 0}$ being the number density of particles in the condensate at equilibrium.

Since we linearize  the kinetic equations, each wavevector component evolves independently. 
Let us define
\begin{equation}
e^{(\pm)}_{{\bf k},{\bf q}}=\frac{{\hbar}^2}{2m}|{\bf k}{\pm}\frac{1}{2}{\bf q}|^2+{\nu}^0-{\mu}.
\end{equation}
The resulting  linearized kinetic equations can be written in the following matrix form,
\begin{eqnarray}
-i{\hbar}\frac{{\partial}{\delta}{{\check{F}}}}{{\partial}t}
={\{} { \epsilon}^{(+)}_{{\bf k},\bf q}~{\delta}{{\check{F}}}-{\delta}{{\check{F}}}~{ \epsilon}^{(-)}_{{\bf k},\bf q}{\}}
+{\hbar}{\bf q}{\cdot}\check{\bf v}_s( {\bf q}){{{F}}}^{eq} -{\hbar}{\bf k}{\cdot}\check{\bf v}_s({\bf q})~{\bf q}{\cdot}{\nabla}_{\bf k}{{{F}}}^{eq}  \nonumber\\
 +{\{}{ B}~{{{F}}}^{eq} 
-  {{{F}}}^{eq} ~{ B'}{\}}  +{\bf q}{\cdot}{\nabla}_{\bf k}{\{}{ D}~{{{F}}}^{eq} 
-{{{F}}}^{eq}~ { D'}{\}} 
+{\delta}{ {\mathcal I}}~~~
\label{partkineq3}
\end{eqnarray}
where 
\begin{equation}
{\delta}{{\check{F}}}=\left( \begin{array}{cc}
{\delta}{\check{F}}_{\it 11}({\bf q},{\bf k},t)&
{\delta}{\check{F}}_{\it 12}({\bf q},{\bf k},t)\\
{\delta}{\check{F}}_{\it 21}({\bf q},{\bf k},t)&
{\delta}{\check{F}}_{\it 22}({\bf q},{\bf k},t)\\
\end{array} \right)  , ~~{ F}^{eq} = \left( \begin{array}{cc}
F^{eq}_{\it 11}({\bf k})&
F^{eq}_{\it 12}({\bf k})\\
F^{eq}_{\it 21}({\bf k})&
F^{eq}_{\it 22}({\bf k})\\
\end{array} \right),
\end{equation}
\begin{equation}
{ \epsilon}^{(+)}_{{\bf k},\bf q}=\left( \begin{array}{cc}
e^{(+)}_{{\bf k},\bf q} &{\Lambda}_0\\
-{\Lambda}_0&- e^{(+)}_{{\bf k},\bf q} \\
\end{array}\right)  , ~~{ \epsilon}^{(-)}_{{\bf k},\bf q}= \left( \begin{array}{cc}
e^{(-)}_{{\bf k},\bf q}& -{\Lambda}_0\\
{\Lambda}_0&- e^{(-)}_{{\bf k},\bf q}\\
\end{array}\right),
\end{equation}
\begin{equation}
{ B}=\left( \begin{array}{cc}
\check{\Psi}(\bf q)&{\delta}{\Lambda}^{\dagger}(\bf q)\\
-{\delta}{\Lambda}(\bf q)&-\check{\Psi}(\bf q)\\
\end{array}\right) , ~~~~{ B'}= \left( \begin{array}{cc}
\check{\Psi}(\bf q)& -{\delta}{\Lambda}^{\dagger}(\bf q)\\
{\delta}{\Lambda}(\bf q)&-\check{\Psi}(\bf q) \\
\end{array}\right),
\end{equation}
\begin{equation}
{ D}=\left( \begin{array}{cc}
-\frac{1}{2}\check{\Psi}(\bf q)&-\frac{1}{2}{\delta}\check{\Lambda}^{\dagger}(\bf q)\\
\frac{1}{2}{\delta}\check{\Lambda}(\bf q)&\frac{1}{2}\check{\Psi}(\bf q)\\
\end{array}\right) , ~~~~{ D'}= \left( \begin{array}{cc}
\frac{1}{2}\check{\Psi}(\bf q)& -\frac{1}{2}{\delta}\check{\Lambda}^{\dagger}(\bf q)\\
\frac{1}{2}{\delta}\check{\Lambda}(\bf q)&-\frac{1}{2}\check{\Psi}(\bf q) \\
\end{array}\right),
\end{equation}
$\check{\Psi}(\bf q)={\hbar} \frac{{\partial}\check{\phi}(\bf q)}{{\partial}t} 
+{\delta}\check{\nu}(\bf q)$  and 
\begin{equation}
{\delta}{ {\mathcal I}}= \left( \begin{array}{cc}
{\delta}  {\mathcal I}_{\it 11}({\bf q},{\bf k},t)&
{\delta}  {\mathcal I}_{\it 12}({\bf q},{\bf k},t)\\
{\delta}  {\mathcal I}_{\it 21}({\bf q},{\bf k},t)&
{\delta}  {\mathcal I}_{\it 22}({\bf q},{\bf k},t)\\
\end{array} \right).
\end{equation}
are the linearized collision integrals for the particle kinetic equations.  
For simplicity and without loss of generality, we have set $v_s^0=0$ (superfluid velocity at equilibrium).

The total particle number density  in the interval  ${\bf q}{\rightarrow}{\bf q}+d{\bf q}$ at time $t$  is 
 is
\begin{equation}
{\delta}{\rm N}({\bf q},t)~=\frac{1}{V}{\sum_{{\bf K}}}{\delta}F_{11}({\bf K},{\bf q},t).
\end{equation}
From Eq. (\ref{partkineq3}) we can write 
\begin{eqnarray}
-i{\hbar}\frac{{\partial}{\delta}F_{11}({\bf k},{\bf q},t)}{{\partial}t}=\left({\tilde \epsilon}^{(+)}_{{\bf k},{\bf q}}-{\tilde \epsilon}^{(-)}_{{\bf k},{\bf q}} \right){\delta}F_{11}({\bf k},{\bf q},t)
+{\hbar}{\bf q}{\cdot}{\bf v}_s( {\bf q},t){\rm N}^{eq}_{\bf k} \nonumber\\
+{\hbar}{\bf q}{\cdot}{\bf v}_s( {\bf q},t)~{\bf q}{\cdot}{\nabla}_{\bf k}{\rm N}^{eq}_{\bf k} -{\Lambda}_0 {\delta}F_{12}({\bf k},{\bf q},t) 
-{\delta}{\Lambda}({\bf q},t)F^{eq} _{12}({\bf k})~~~~~~~~~~ \nonumber\\
+{\Lambda}_0{\delta}F_{21}({\bf k},{\bf q},t)+{\delta}{\Lambda}^{\dagger}({\bf q},t)F^{eq} _{21}({\bf k})+{\delta}{\mathcal I}_{11}({\bf k},{\bf q},t)~~~~~~~~
\label{parkineq2.aaa}
\end{eqnarray}
 Note that 
\begin{eqnarray}
{\Lambda}_0=\frac{g}{V}{\sum_{{\bf k}}}F^{eq} _{12}({\bf k})=\frac{g}{V}{\sum_{{\bf k}}}F^{eq} _{12}({\bf k}),~~~~~~~~~~~~~\nonumber\\
{\delta}{\Lambda}({\bf q},t)=\frac{g}{V}{\sum_{{\bf k}}}{\delta}F_{21}({\bf k},{\bf q},t),~~~{\delta}{\Lambda}^{\dagger}({\bf q},t)=\frac{g}{V}{\sum_{{\bf k}}}{\delta}F_{12}({\bf k},{\bf q},t).
\end{eqnarray}
Let us now sum over all momentum states in Eq. (\ref{parkineq2.aaa}). The terms that depend on ${\Delta}$ cancel. One can also check that the third term on the right hand side of Eq. (\ref{parkineq2.aaa}) gives a negligible contribution, compared to the second term, when one integrates over ${\bf k}$. 
Eq. (\ref{parkineq2.aaa})  then reduces to
\begin{eqnarray}
-i{\hbar}\frac{{\partial}{\delta}{\rm N}({\bf q},t)}{{\partial}t}=\frac{ {\hbar}^2}{m}\frac{1}{V}{\sum_{{\bf k}}}{\bf k}{\cdot}{\bf q}{\delta}F_{11}({\bf k}, {\bf q},t) 
+{\hbar}{\bf q}{\cdot}{\bf v}_s( {\bf q},t){\rm N}^{eq} ~~~~~~~~
\label{conteq}
\end{eqnarray}
which is the continuity equation for total particle number density.

\subsection{Bogolon Kinetic Equation}

The dynamics of the  excitations (bogolons) in the BEC governs the hydrodynamic relaxation of the BEC.  In order to determine the hydrodynamic behavior of the BEC, we must transform from the particle kinetic equations to kinetic equations for the  bogolons.

We can transform from the particle kinetic equation to the bogolon kinetic equation using the Bogoliubov transformation ${ S}_j$ (see Appendix C) which transforms particle creation and annihilation operators, 
 ${\hat a}^{\dagger}_j={\hat a}^{\dagger}_{{\bf k}_j}$ and ${\hat a}_j={\hat a}_{{\bf k}_j}$ respectively,  into bogolon creation and annihilation operators, ${\hat b}^{\dagger}_j={\hat b}^{\dagger}_{{\bf k}_j}$ and ${\hat b}_j={\hat b}_{{\bf k}_j}$, respectively.
 We can write 
\begin{equation}
\left( \begin{array}{cc}
{\langle}{\hat  b}^{\dagger}_{1}{\hat  b}_{2} \rangle &
0\\
0&
{\langle} {\hat  b}_{-1}{\hat  b}^{\dagger}_{-2} \rangle \\
\end{array} \right)={ S}^{-1}_1{\cdot}~\left( \begin{array}{cc}
{\langle}{\hat  a}^{\dagger}_{1}{\hat  a}_{2}  \rangle  &
{\langle}{\hat  a}^{\dagger}_{1}{\hat  a}^{\dagger}_{-2}  \rangle  \\
{\langle} {\hat  a}_{-1}{\hat  a}_{2}  \rangle  &
{\langle} {\hat  a}_{-1}{\hat  a}^{\dagger}_{-2}  \rangle  \\
\end{array} \right)  {\cdot} { S}^{-1}_2,
\end{equation}
where
\begin{equation}
{ S}_j=\left( \begin{array}{cc}
u_j&
-v_j\\
-v_j&
u_j\\
\end{array} \right)~~~{\rm and}~~~{ S}^{-1}_j=\left( \begin{array}{cc}
u_j&
v_j\\
v_j&
u_j\\
\end{array} \right), \ \ j=1,2
\end{equation}
with $u_1=u_{\kk_1}$, $u_2=u_{\kk_2}$ (see Appendix B). 
Since  excitations (the bogolons) do not form a condensate, we require that ${\langle}{\hat  b}^{\dagger}_{1}{\hat  b}^{\dagger}_{-1}\rangle=0$ and 
${\langle} {\hat  b}_{-1}{\hat  b}_{2}{\rangle}=0$.  Also, since we are linearizing the kinetic equations about absolute equilibrium,  we can express the  parameters $u_1$ and $v_1$ in terms of equilibrium quantities. 
We also find 
\begin{equation}\begin{aligned}
 &\ \left( \begin{array}{cc}
{\mathcal N}^{eq}({\bf k})&0\\
0&{\mathcal N}^{eq}({\bf k})+1\\
\end{array} \right)= \left( \begin{array}{cc}
{\langle}{\hat b}_{\bf k}^{\dagger}{\hat b}_{\bf k}{\rangle}_{eq}&0\\
0&
{\langle}{\hat b}_{-{\bf k}}{\hat b}^{\dagger}_{-{\bf k}}{\rangle}_{eq}\\
\end{array} \right)
={S}_{\bf k}^{-1}{\cdot}{ F}^{eq}({\bf k}){\cdot}{S}_{\bf k}^{-1},\end{aligned}
\label{partlin1}
\end{equation}
where ${\mathcal N}^{eq}_{\bf k}=\left[{\rm exp}({\beta}E_{\bf k})-1\right]^{-1}$ is the Bose-Einstein distribution for bogolons.

We can expand the particle number distribution in terms of bogolon distributions to obtain
\begin{equation}
{\delta}\check{F}_{11}({\bf q},{\bf k},t)
= u^2_{{k}}f({\bf q},{\bf k},t)+v^2_{{k}}f({\bf q},-{\bf k},t).
\label{phieq7}
\end{equation}
and  expand the particle current in terms of bogolon currents to obtain 
\begin{eqnarray}
{\sum_{{\bf k}}} {\bf k}{\delta}F_{11}({\bf q},{\bf k},t)={\sum_{{\bf k}}} {\bf k} {\big[} u^2_{k}f({\bf q},{\bf k},t)
+v^2_{k}f({\bf q},-{\bf k},t) {\bigr]}  ~~\nonumber\\
={\sum_{{\bf k}}} {\bf k}{\delta}f({\bf q},{\bf k},t).~~~~~~~~~~~~~~~~~~~~
\label{phieq8}
\end{eqnarray}
since $u_{k}^2-v_{k}^2=1$.  Thus, we find that the bogolon momentum density is equal to the particle momentum density.

We now  write  the Hugenholtz-Pines (H-P) equation, ${\mu}={\nu}-{\Delta}$ for the BEC \cite{hugenholtz}. Since the time derivative of the macroscopic phase ${\phi}({\bf q},t)$ is proportional to the chemical potential ${\mu}={\hbar} \frac{{\partial}{\phi}}{{\partial}t}$, in  the hydrodynamic regime, where  we can assume that the system is locally in equilibrium, and write
\begin{equation}
{\hbar} \frac{{\partial}{\phi}({\bf q},t)}{{\partial}t} 
+{\delta}{\nu}({\bf q},t)-{\delta}{\tilde \Lambda}({\bf q},t)=0,
\label{H-P2}
\end{equation}
where 
\begin{equation}
{\delta}{\nu}({\bf q},t)=2g{\sum_{\bf K}}{\delta}F_{11}({\bf q},{\bf k})=2g{\delta}{\rm N}({\bf q},t),
\end{equation}
and 
\begin{equation}
{\delta}{\tilde \Lambda}({\bf q},t)=\frac{g}{2}{\sum_{\bf k}}({\delta}F_{12}({\bf q},{\bf k},t)+{\delta}F_{21}({\bf q},{\bf k},t)).
\end{equation}
We now make a``Bogoliubov-like" approximation for the nonequilibrium order parameter,  ${\delta}{\tilde \Lambda}({\bf q},t)=g{\delta}{\rm N}({\bf q},t)$. Then the Hugenholtz-Pines equation can  be written in the form
\begin{equation}
{\hbar} \frac{{\partial}{\phi}({\bf q},t)}{{\partial}t} 
+g{\delta}{\rm N}({\bf q},t)=0.
\label{H-P2a}
\end{equation}
This approximation limits us to very dilute gases, and we expect that it limits the accuracy of our results for the longitudinal modes to the temperature range $0{\leq}T{\leq}0.3T_c$.
Equation (\ref{H-P2a}) gives a closure condition for the hydrodynamic equations. Using Eqs. (\ref{conteq}),  (\ref{phieq8}), and (\ref{H-P2a}),
we obtain the system of equations in Eqs.  \eqref{kineq1a}-\eqref{kineq1b}-\eqref{phi1a}-\eqref{phi1b}.

\section{Microscopic hydrodynamic modes }

In order to obtain the dispersion relation for the hydrodynamic  modes  of the BEC,  we consider one frequency component of the linearized kinetic equations. We can  write
\begin{equation}
f{(\bf q},\kk,t)~{\sim}~{\rm e}^{i{\omega}t}~{\tilde {\rm f}}(\bf q, \kk,\omega) ~~~{\rm and}~~~\phi({\bf q},t)~{\sim}~{\rm e}^{i{\omega}t}~{\tilde \varphi}(\bf q,\omega).
\label{fourier}
\end{equation}
Then Eq. (\ref{kineq1a}) takes the form
\begin{equation}
{\omega}{\tilde {\rm f}}({\bf q},\kk,\omega)=\frac{{\hbar}}{m}{\bf k}{\cdot}{\bf q}\frac{{\epsilon}(\kk)+\Lambda_0}{E_{\kk}} ~{\tilde {\rm f}}({\bf q},\kk,\omega)
  -i \frac{{\hbar}}{m}~q^2{\tilde  {\varphi}}({\bf q},{\omega}){\mathcal N}^{\rm eq}_{{\bf k}}+i{\bf G}{[}{\tilde {\rm f}}{]}(\bf q,\kk,\omega),
\label{kineq4}
\end{equation}
where we have used the fact that  ${\bf v}_s({\bf q},\omega)=-i   \frac{\hbar}{m}{\bf q} {\tilde {\varphi}}({\bf q},\omega)$.  Equation (\ref{phi1a}) takes the form
\begin{eqnarray}
{\omega}^2 {\tilde {\varphi}}({\bf q},\omega)
=i\frac{g}{m}~\frac{1}{(2{\pi})^3}{\int_{\bf{R}^3}}d{\bf k}~{\bf q}{\cdot}{\bf k}~{\tilde {\rm f}}({\bf q},\kk,\omega) +q^2 \frac{g}{m} {\tilde {\varphi}}({\xi},\omega){n}^{eq}.
\label{phase4}
\end{eqnarray}
Equations (\ref{kineq1a})-(\ref{kineq1b}) and (\ref{phi1a})-(\ref{phi1b}) are the bogolon kinetic equations that describe hydrodynamic behavior of a dilute BEC.

We can combine Eqs. (\ref{kineq4}) and (\ref{phase4}) and obtain
\begin{equation}
\begin{split}
{\omega}~{\tilde {\rm f}}({\bf k},{\bf q},{\omega}) - \frac{q^2}{{\omega}^2- v_B^2 q^2} \frac{g\hbar}{m^2}{\N}_k^{\rm eq}
~\frac{1}{(2\pi)^3} {\int}d{{\bf k}_1}~{\bf q}{\cdot}{\bf k}_1~{\tilde {\rm f}}({\bf k}_1,{\bf q},{\omega}) 
\nonumber\\
= {\bf q}{\cdot}{\bf k}~\frac{{\hbar}}{m}\frac{({\epsilon}_k+{\Lambda}_0)}{E_k}{\tilde {\rm f}}({\bf k},{\bf q},{\omega}) 
+i{\bf G}{[}{\tilde {\rm f}}{]}(\bf q,\kk,\omega)
\label{totkin1}
\end{split}
\end{equation}
where $v_B = \sqrt{\frac{g n^{\rm eq}}{m}} $ is the Bogoliubov speed. 
Let us define
\begin{equation}
{\tilde {\rm f}}({\bf q}\kk,{\omega})={\N}^{\rm eq}_{\kk}{\mathcal M}^{\rm eq}_{\kk}h({\bf q},\kk,{\omega}),
\label{small}
\end{equation}
where ${\mathcal M}^{\rm eq}_{\kk}=(1+{\N}^{\rm eq}_{\kk}) $. Then Eq. (\ref{totkin1}) takes the form
\begin{eqnarray}
{\omega}~h({\bf k},{\bf q},{\omega}) - \frac{q^2}{{\omega}^2- v_B^2 q^2} \frac{g\hbar}{m^2}\frac{1}{{\mathcal M}_{\bf k}^{\rm eq}}
~\frac{1}{(2\pi)^3} {\int}d{{\bf k}_1}~{\bf q}{\cdot}{\bf k}_1~{\N}^{\rm eq}_{\kk_1}{\mathcal M}_{{\bf k}_1}^{\rm eq} h({\bf q},\kk_1,{\omega})
~~~~~~\nonumber\\
= {\bf q}{\cdot}{\bf k}~\frac{{\hbar}}{m}\frac{({\epsilon}_k+{\Lambda}_0)}{E_k}h({\bf k},{\bf q},{\omega}) ~~~~~~~~~~~~~~~~~~~~~~~~~~~~~~~~~~~~~~~~~~~~~~~~\nonumber\\
+ i~{\int_0^{\infty}}dk_1{\int}d{\Omega}_1~\sqrt{\frac{k_1^2{\N}_{k_1}^{\rm eq}{\mathcal M}_{k_1}^{\rm eq}}{k^2{\N}_k^{\rm eq}{\mathcal M}_k^{\rm eq}}}~{\mathcal C}({\bf k},{\bf k}_1)  h({\bf k}_1,{\bf q},{\omega}),~~~~~~~~~~~~~~~
\label{totkin2}
\end{eqnarray}
where ${\mathcal C}({\bf k},{\bf k}_1)$ is the bogolon collision operator whose properties are described in Appendix C. Eq. (\ref{totkin2}) is an eigenvalue equation with an unusual structure.  This becomes clearer if we write
\begin{equation}
h({\bf k},{\bf q},{\omega}) =\frac{1}{\sqrt{k^2{\N}_{k}^{\rm eq}{\mathcal M}_{k}^{\rm eq}}} {\Upsilon}({\bf q},{\bf k},{\omega})
\label{ eig1}
\end{equation}
Then the eigenvalue equation takes the form
\begin{eqnarray}
{\omega}~{\Upsilon}({\bf q},{\bf k},{\omega})- \frac{q^2}{{\omega}^2- v_B^2 q^2} \frac{g\hbar k}{m^2}\sqrt{\frac{{\N}_{\bf k}^{\rm eq}}{{\mathcal M}_{\bf k}^{\rm eq}}}
~\frac{1}{(2\pi)^3} {\int}d{{\bf k}_1}\frac{1}{k_1}~{\bf q}{\cdot}{\bf k}_1~\sqrt{{\N}^{\rm eq}_{\kk_1}{\mathcal M}_{{\bf k}_1}^{\rm eq}} {\Upsilon}({\bf q},{\bf k}_1,{\omega})
\nonumber\\
= {\bf q}{\cdot}{\bf k}~\frac{{\hbar}}{m}\frac{({\epsilon}_k+{\Lambda}_0)}{E_k} {\Upsilon}({\bf q},{\bf k},{\omega})
+ i~{\int_0^{\infty}}dk_1{\int}d{\Omega}_1~~{\mathcal C}({\bf k},{\bf k}_1)  {\Upsilon}({\bf q},{\bf k},{\omega}).~~~~~~~~~
\label{totkin3}
\end{eqnarray}
The hydrodynamic behavior occurs for long wavelength (small $q$) processes.  Therefore, it is enough to consider Eq. (\ref{totkin3}) for small $q$.  Without loss of generality, we can assume that ${\bf q}=q{\hat e}_z$, where ${\hat e}_z$ is the unit vector along the z-direction.

We will use perturbation theory to solve Eq. (\ref{totkin3})  to second order in $q$. We expand 

\begin{eqnarray}
{\omega}={\omega}^{(0)}+q{\omega}^{1)}+q^2{\omega}^{(2)}+...~~~~~~~~\nonumber\\
{\Upsilon}({\bf q},{\bf k})={\Upsilon}^{(0)}({\bf k})+q{\Upsilon}^{(1)}({\bf k})+q^2{\Upsilon}^{(2)}({\bf k})+...
\end{eqnarray}

The time dependence of  hydrodynamic  modes has the form 
${\rm e}^{ i({\omega}^{1)}q+{\omega}^{(2)}q^2+...)t}$. If the  frequency ${\omega}^{(1)}$ is non-zero, the mode is propagating and  ${\omega}^{(1)}$ is the speed of propagation of the mode.  The  frequency ${\omega}^{(2)}$  gives the decay rate of the mode.  The non-hydrodynamic modes have a zeroth order contribution ${\omega}^{(0)}$ which generally causes them to decay rapidly.

\subsubsection{Zeroth Order Perturbation Theory}

For $q=0$, Eq. (\ref{totkin3})  reduces to
\begin{equation}
\begin{split}
{\omega}^{(0)}~{\Upsilon}^{(0)}({\bf k})=
+ i~{\int_0^{\infty}}dk_1{\int}d{\Omega}_1~~{\mathcal C}({\bf k},{\bf k}_1)  {\Upsilon}^{(0)}({\bf k}).
\label{totord0}
\end{split}
\end{equation}
where ${\Upsilon}^{(0)}({\bf k}){\equiv}{\Upsilon}({\bf 0},{\bf k})$. Thus, to zeroth order in $q$, ${\Upsilon}^{(0)}({\bf k})$ is an eigenvector of the collision operator with eigenvalue equal to ${\omega}^{(0)}$.  As discussed in Appendix C, the collision operator has four eigenvalues equal to zero with eigenfunctions that depend on bogolon  energy and momentum,  which are conserved during the collisions between bogolons.  

It is useful to note that the bogolon collision operator differs from that of a monatomic gas of classical particles,   which would have a fifth zero eigenvalue corresponding to conservation of particle number during collisions.  The remaining nonzero eigenvalues of the collision operator are negative.  
For a classical gas, the five zero eigenvalues are the source of the  five hydrodynamic modes of the gas.  In the  BEC, there are only four zero eigenvalues, but we know that the BEC has six hydrodynamic modes. The remaining two hydrodynamic modes in the BEC come from the nonlinear dependence on ${\omega}$  in the eigenvalue equation (\ref{totkin3}).  This, in turn comes from the coupling of the macroscopic phase to the bogolon kinetic equation.  A classical monatomic gas has one pair of propagating hydrodynamic modes which are sound modes.  As we will see, a BEC has two pairs of propagating modes, corresponding to first and second sound.  The additional pair of  sound modes in a BEC comes from coupling of the bogolon kinetic equation to the macroscopic phase.

\subsubsection{First Order Perturbation Theory}

Since the eigenvalues of the collision operator are four-fold degenerate at zeroth order, it is necessary to find the correct combination of zeroth order eigenstates, in order to  compute the contributions first order and second order in $q$.  

To first order in $q$, Eq. (\ref{totkin3}) can be written 
\begin{eqnarray}
{\omega}^{(1)}~{\Upsilon}^{(0)}({\bf k})- \frac{1}{({\omega}^{(1)})^2- v_B^2 } \frac{g\hbar k}{m^2}\sqrt{\frac{{\N}_{\bf k}^{\rm eq}}{{\mathcal M}_{\bf k}^{\rm eq}}}
~\frac{1}{(2\pi)^3} {\int}d{{\bf k}_1}\frac{1}{k_1}~k_{1,z}~\sqrt{{\N}^{\rm eq}_{\kk_1}{\mathcal M}_{{\bf k}_1}^{\rm eq}} {\Upsilon}^{(0)}({\bf k}_1)
\nonumber\\
= k_z~\frac{{\hbar}}{m}\frac{({\epsilon}_k+{\Lambda}_0)}{E_k} {\Upsilon}^{(0}({\bf k})
+ i~{\int_0^{\infty}}dk_1{\int}d{\Omega}_1~~{\mathcal C}({\bf k},{\bf k}_1)  {\Upsilon}^{(1)}({\bf k}).
~~~~~~~~
\label{totord1}
\end{eqnarray}
Equation (\ref{totord1}) is not symmetric with respect to operations from the left and right Therefore, we must find different combinations of eigenstates of ${\mathcal C}({\bf k},{\bf k}_1) $ for operations to the right and to the left of Eq(\ref{totord1}). Let us write
\begin{eqnarray}
{\Upsilon}_R^{(0)}({\bf k})={\sum_{{\ell}=0}^{1}}{\sum_{m=-{\ell}}^{\ell}}~{\Gamma}_{0,{\ell},m}^R{\Upsilon}^{(0)}_{0,\ell,m}({\bf k})~{\rm and}~{\Upsilon}_L^{(0)}({\bf k})={\sum_{{\ell}=0}^{1}}{\sum_{m=-{\ell}}^{\ell}}~{\Gamma}_{0,{\ell},m}^L{\Upsilon}^{(0)}_{0,\ell,m}({\bf k}),~\nonumber\\
\end{eqnarray}
where ${\Upsilon}_L^{(0)}({\bf k})$ and ${\Upsilon}_R^{(0)}({\bf k})$ are left and right  eigenvectors respectively,  of ${\mathcal C}({\bf k},{\bf k}_1) $ with eigenvalue zero.  When we multiply Eq. (\ref{totord1}) from the left with  ${\Upsilon}_L^{(0)}({\bf k})$, the contribution from the collision operator drops out, and we finally obtain the matrix equation
${\bar {\Gamma}}^{\dagger}_L{\cdot}{\bar M}{\cdot}{\bar {\Gamma}}_R=0$, where ${\bar {\Gamma}}^{\dagger}_R$ is the row matrix 
\[
{\bar {\Gamma}}^{\dagger}_L={\{}{\Gamma}_{0,0,0}^{L*},{\Gamma}_{0,1,0}^{L*},{\Gamma}_{01,1}^{L*},{\Gamma}_{0,1,-1}^{L*}{\}}
\]
 and ${\bar {\Gamma}}_R$ is a column matrix whose transpose is 
 \[
 {\bar {\Gamma}}^T_R={\{}{\Gamma}_{0,0,0}^{R},{\Gamma}_{0,1,0}^{R},{\Gamma}_{01,1}^{R},{\Gamma}_{0,1,-1}^{R}{\}}
 \] 
The matrix ${\bar M}$ is given by
\begin{eqnarray}
{\bar M}=\left(  \begin{array}{cccc}
-{\omega}^{(1)}&{\alpha}+\frac{\gamma}{({\omega}^{(1)})^2-v_B^2}&0&0\\
{\alpha}&-{\omega}^{(1)}&&0\\
0 & 0 & -{\omega}^{(1)}&0\\
0 & 0 & 0 &-{\omega}^{(1)}\\
\end{array}\right)
\end{eqnarray}
where 
\begin{eqnarray}
{\alpha}={\int}d{\bf k}{\Upsilon}^{(0)*}_{0,0,0}({\bf k}) k_z~\frac{{\hbar}}{m}\frac{({\epsilon}_k+{\Lambda}_0)}{E_k} {\Upsilon}^{(0)}_{0,0,1}({\bf k})
\end{eqnarray}
and
\begin{eqnarray}
{\gamma}= \frac{g\hbar}{m^2(2\pi)^3}{\int}d{\bf k}k{\Upsilon}^{(0)*}_{0,0,0}({\bf k}) \sqrt{\frac{{\N}_{\bf k}^{\rm eq}}{{ \mathcal M }_{\bf k}^{\rm eq}}}
~ {\int}d{{\bf k}_1}\frac{k_{1,z}}{k_1}~\sqrt{{\N}^{\rm eq}_{\kk_1}{\mathcal M }_{{\bf k}_1}^{\rm eq}} {\Upsilon}^{(0)}_{0,0,1}({\bf k}_1)\end{eqnarray}

The values of ${\omega}^{(1)}$ can be found from the condition $Det[{\bar M}]=0$. 
This gives solutions
\begin{eqnarray}
{\omega}^{(1)}_2=-{\omega}^{(1)}_1=\frac{1}{\sqrt{2}}\sqrt{v_B^2+{\alpha}^2-\sqrt{(v_B^2-{\alpha}^2)^2+4{\alpha}{\gamma}}}~
~~\nonumber\\
{\omega}^{(1)}_4=-{\omega}^{(1)}_3=\frac{1}{\sqrt{2}}\sqrt{v_B^2+{\alpha}^2+\sqrt{(v_B^2-{\alpha}^2)^2+4{\alpha}{\gamma}}}~~
~~\nonumber\\
{\omega}^{(1)}_6={\omega}^{(1)}_5=0.~~~~~~~~~~~~~~~~~~~~~~~~~
\end{eqnarray}
Thus, at first order there are six hydrodynamic frequencies. The frequencies ${\omega}^{(1)}_2=-{\omega}^{(1)}_1$ correspond to fast sound modes.  The frequencies ${\omega}^{(1)}_4=-{\omega}^{(1)}_3$ correspond to fast sound modes. The frequencies ${\omega}^{(1)}_6=-{\omega}^{(1)}_5=0$ correspond to non-propagating transverse viscous modes in the BEC.  

In Fig. 1, we plot the sound speed, in units of the Bogoliubov speed $v_B$,   as a function of fractional distance below the critical temperature for $n^{eq}a^3=10^{-6}$ \cite{ReichlGust:2014:DOH}, which is a  value found in experiments.
For a  rubidium  BEC, it would correspond to a  scattering length $a=5.6{\times}10^{-9}$m and a density $n^{eq}=5.7{\times}10^{18}{\rm m}^{-3}$.  There are several things to note. (i) The speed of both the fast mode and the slow mode approach finite values in the limit $T{\rightarrow}0$ K. This behavior of the sound speeds is consistent with the behavior of the sound speeds found by Lee and Yang  \cite{lee1956question} using a very different approach.  It is a consequence of the fact that the bogolon spectrum becomes phonon-like at very low temperature.  One does not see this behavior of the sound speeds for models with particle-like spectrum at very low temperature \cite{griffin1997first}. For models with particle-like spectrum, one of the sound speeds goes to zero as $T{\rightarrow}0$ K.  (ii) The sound speeds undergo an avoided crossing as the temperature is lowered.  The temperature at which the avoided crossing occurs increases with increasing density of the gas.  For $n^{eq}a^3=10^{-5}$ it occurs at $T/T_c{\approx}0.05$. For $n^{eq}a^3=10^{-4}$ it occurs at $T/T_c{\approx}0.11$  \cite{ ReichlGust:2014:DOH}.

The zeroth order left and right eigenstates can be obtained from the equations ${\bar {\Gamma}}^{\dagger}_L{\cdot}{\bar M}=0$, and ${\bar M}{\cdot}{\bar {\Gamma}}_R=0$, respectively, once the frequencies are determined.  For the transverse modes, the left and right eigenstates are complex conjugates of one another and  ${\Upsilon}_R^{(0)}({\bf k})={\Upsilon}_{0,1,{\pm}1}^{(0)}({\bf k})$.  For the longitudinal modes, the left and right eigenstates are of the form 
${\Upsilon}_R^{(0)}({\bf k})={\Gamma}_{0,0,0}{\Upsilon}_{0,0,0}^{(0)}({\bf k})+{\Gamma}_{0,1,0}{\Upsilon}_{0,1,0}^{(0)}({\bf k})$.  The exact expression for ${\Gamma}_{0,0,0}$ and ${\Gamma}_{0,0,0}$ for a given mode can  be found by solving ${\bar M}{\cdot}{\bar {\Gamma}}_R=0$ for the particular first order frequency considered.

\subsubsection{Second Order Perturbation Theory}

To second order in $q$, Eq. (\ref{totkin3}) takes the form
\begin{eqnarray}
{\omega}^{(2)}~{\Upsilon}^{(0)}({\bf k})+{\omega}^{(1)}~{\Upsilon}^{(1)}({\bf k})~~~~~~~~~~~~~~~~~
 \nonumber\\
- \frac{1}{({\omega}^{(1)})^2- v_B^2 } \frac{g\hbar k}{m^2}\sqrt{\frac{{\N}_{\bf k}^{\rm eq}}{{ \mathcal M }_{\bf k}^{\rm eq}}}
~\frac{1}{(2\pi)^3} {\int}d{{\bf k}_1}\frac{k_{1,z}}{k_1}~\sqrt{{\N}^{\rm eq}_{\kk_1}{ \mathcal M}_{{\bf k}_1}^{\rm eq}} {\Upsilon}^{(1)}({\bf k}_1)
\nonumber\\
+\frac{2{\omega}^{(1)}{\omega}^{(2)}}{({\omega}^{(1)})^2- v_B^2)^2 } \frac{g\hbar k}{m^2}\sqrt{\frac{{\N}_{\bf k}^{\rm eq}}{{ \mathcal M}_{\bf k}^{\rm eq}}}
~\frac{1}{(2\pi)^3} {\int}d{{\bf k}_1}\frac{k_{1,z}}{k_1}~\sqrt{{\N}^{\rm eq}_{\kk_1}{ \mathcal M}_{{\bf k}_1}^{\rm eq}} {\Upsilon}^{(0)}({\bf k}_1)
\nonumber\\
= k_z~\frac{{\hbar}}{m}\frac{({\epsilon}_k+{\Lambda}_0)}{E_k} {\Upsilon}^{(1)}({\bf k})
+ i~{\int_0^{\infty}}dk_1{\int}d{\Omega}_1~~{\mathcal C}({\bf k},{\bf k}_1)  {\Upsilon}^{(2)}({\bf k}_1).
\label{totord2}
\end{eqnarray}
The state ${\Upsilon}^{(1)}({\bf k})~$ can be obtained from Eq. (\ref{totord1}). If we multiply on the left by $ {\Upsilon}_L^{(0)}({\bf k})$ and integrate, we can eliminate the collision operator from this equation and obtain values for ${\omega}^{(2)}$ for each of the hydrodynamic modes.  These quantities are the decay rates of the modes.

In order to write explicit expressions for the decay rates, it is useful to introduce  abstract notation.  All the eigenfunctions of the collision operator 
 $C({\bf k}_1,{\bf k}_2)$ can be written in the form
\begin{equation}
{\Upsilon}_{\beta,\ell,m}({\bf k})={\Upsilon}_{\beta,\ell}(k)~{\rm Y}_{\ell}^m({\hat {\bf k}}),
\end{equation}
where ${\rm Y}_{\ell}^m({\hat {\bf k}})$ is a spherical harmonic and  ${\hat {\bf k}}={\bf k}/|{\bf k}|$. The eigenstates are orthonormalized so that
\begin{equation}
{\int_0^{\infty}}dk_1{\int}d{\Omega}_1~{\Upsilon}^*_{\beta_1,\ell_1,m_1}({\bf k}_1){\Upsilon}_{\beta_2,\ell_2,m_2}({\bf k}_1)={\delta}_{\beta_1,\beta_2}{\delta}_{\ell_1,\ell_2}{\delta}_{m_1,m_2}
\end{equation}
and ${\int_0^{\infty}}dk_1~{\Upsilon}^*_{\beta_1,\ell}(k_1){\Upsilon}_{\beta_2,\ell}(k_1)={\delta}_{\beta_1,\beta_2}$.  We can also write the collision operator in a spectral decomposition
\begin{eqnarray}
C({\bf k}_1,{\bf k}_2)={\sum_{\ell=0}^{\infty}}{\sum_{m=-\ell}^{\ell}}~{\mathcal C}_{\ell}(k_1,k_2)~{\rm Y}_{\ell}^{m}({\hat {\bf k}}_1){\rm Y}_{\ell}^{m*}({\hat {\bf k}}_2)~~~~~~\nonumber\\
={\sum_{\beta=0}^{\infty}}{\sum_{\ell=0}^{\infty}}{\sum_{m=-\ell}^{\ell}}~{\lambda}_{\beta,\ell}~{\Upsilon}_{\beta,\ell,m}({\bf k}_1){\Upsilon}^*_{\beta,\ell,m}({\bf k}_2)
\label{CYseries1}
\end{eqnarray}
where ${\mathcal C}_{\ell}(k_1,k_2)={\sum_{\beta=0}^{\infty}}~{\lambda}_{\beta,\ell}~{\Upsilon}_{\beta,\ell}(k_1){\Upsilon}^*_{\beta,\ell}(k_2)~$. Here  ${\Upsilon}_{\beta,\ell,m}({\bf k}_1)$ (with ${\beta}{\neq}0$)  are eigenstates of  ${\mathcal C}_{\ell}(k_1,k_2)$ with eigenvalues ${\lambda}_{\beta,\ell}<0$.

We can now  express the operator ${\mathcal C}_{\ell}(k_1,k_2)$ in ``bra-ket" notation as
\begin{equation}
{\hat {\mathcal C}}_{\ell}={\sum_{\beta=0}^{\infty}}{\lambda}_{\beta,\ell}|{\Upsilon}_{\beta,\ell}{\rangle}{\langle}{\Upsilon}_{\beta,\ell}|,
\end{equation}
so that ${\langle}k_1|{\hat {\mathcal C}}_{\ell}|k_2{\rangle}={\mathcal C}_{\ell}(k_1,k_2)$ and ${\langle}k|{\Upsilon}_{\beta,\ell}{\rangle}={\Upsilon}_{\beta,\ell}(k)$,  where ${\langle}k_1|k_2{\rangle}={\delta}(k_1-k_2)$ and ${\int_0^{\infty}}dk~ |k{\rangle}{\langle}k|={\hat 1}$, where ${\hat 1}$ is the unit operator.

The decay rate for the transverse (viscous) modes can be written
\begin{equation}
{\omega}^{(2)}_6=\frac{1}{5} {\sum_{ {\beta}=0}^{\infty}} \frac{1}{{\lambda}_{\beta,2} } {\big|} {\langle} {\Upsilon}_{\beta,\ell} | kB_k|{\Upsilon}_{\beta,2} {\rangle} {\big|}^2
\end{equation}
where
\begin{equation}
B_k= \frac{{\hbar}}{m} \frac{ {\epsilon}_k+{\Lambda}_0}{E_k}
\end{equation}
The viscosity ${\eta}$ of the BEC is related to ${\omega}^{(2)}_6$ via the equation ${\eta}={\rho}_n{\omega}^{(2)}_6$, where ${\rho}_n$ is the density of the normal (non-condensate) part of the the BEC \cite{gust2015viscosity}.

The decay rates for the longitudinal (sound) modes are given by
\begin{eqnarray}
{\omega}^{(2)} =\frac{i}{1+{\mathcal S}}~{\bigl{\{}}\frac{1}{6}{\rm C}_{0;1}+\frac{1}{6}{\rm C}_{1;0}+\frac{2}{15}{\rm C}_{1;2}+C^{'}_0{\bigr{\}}},
\label{omega2f1}
\end{eqnarray}
where
\begin{equation}
{\rm C}_{{\ell}',{\ell}}={\sum_{{\beta}=0}^{\infty}}   \frac{1}{{\lambda}_{{\beta},\ell}} |{\langle}{\Upsilon}_{0,{\ell}'}|kB_k| {\Upsilon}_{\beta,{\ell}} {\rangle}|^2,
\label{S1}
\end{equation}
\begin{eqnarray}
C^{'}_0=\frac{g{\hbar}}{12{\pi}^2m^2D_{0,1}}\frac{1}{(({\omega}^{(1)})^2-v_B^2)}~~~~~~~~~~~~~~~~~~~~~\nonumber\\
{\times}{\sum_{{\beta}=0}^{\infty}}   \frac{1}{{\lambda}_{{\beta},0}} {\langle}{\Upsilon}_{0,{\ell}'}|kB_k|{\Upsilon}_{{\beta},0}{\rangle}{\int}dk k {\psi}^{*}_{\beta,0}(k)\sqrt{\frac{{\N}_{\bf k}^{\rm eq}}{{ \mathcal M }_{\bf k}^{\rm eq}}}
\end{eqnarray}

and
\begin{equation}
{\mathcal S}=\left[\frac{({\omega}^{(1)})^2}{(({\omega}^{(1)})^2-v_B^2)^2}
\right] \frac{g\hbar}{m^2{\alpha}}
~~\frac{1}{\sqrt{3}}\frac{1}{2{\pi}^2}~\frac{1}{D_{0,1}}~{\int}dk k\sqrt{\frac{{\N}_{\bf k}^{\rm eq}}{{  \mathcal M}_{\bf k}^{\rm eq}}} {\psi}_{0,0}(k).
\label{S1}
\end{equation}
The lifetime of the sound modes in the BEC is given by $({\omega}^{(2)}q^2)^{-1}$ and  depends on the speed ${\omega}^{(1)}$ of the sound mode. On the right hand side of (\ref{omega2f1}), the first three terms  are current-current correlation functions similar to those that determine the decay of sound modes in classical gases.  The factor of ${\mathcal S}$ in the denominator is a consequence of the macroscopic phase that results from the broken gauge symmetry in the BEC below $T=T_c$.

\subsection{Comparison to Experiment}

We can compare the  prediction of Eq. (\ref{omega2f1}), for the decay rate of sound modes in BECs, to the results found in the Steinhauer experiment \cite{steinhauer12}  in which a sound mode was excited  in  a $^{87}{\rm Rb}$ BEC, and observed to decay. The  wavelength of the sound wave was about $18{\times}10^{-6}~{\rm m} ~(q = 0.35 \mu {\rm m}^{-1})$.  
The particle  density of the BEC was about $n^{\rm eq}=9.71{\times}10^{19}~{\rm m}^{-3}$ which gives a critical temperature of about $T_c=3.90{\times}10^{-7}~{\rm K}$. The Bogoliubov speed in this case is approximately   $v_B \approx 1.887 {\rm mm} / {\rm s}$, which is  approximately the sound speed observed in the experiment \cite{steinhauer12}. In the experiment, a harmonic trap with frequencies ${f}_1={ f}_2=224~{\rm Hz}$ and ${f}_3=26~{\rm Hz}$ was used to create a 1D sound mode.  Around ${\langle}N{\rangle}=5{\times}10^5$ atoms in the trap were used, which gives the  critical temperature $T_C=\frac{h}{m}\left(\frac{{\langle}N{\rangle}{f}_1{f}_2{f}_3}{1.202}\right)^{1/3}{\approx}3.9{\times}10^{-7}$.  The sound mode  had a wave vector $q=0.35~{\mu}{\rm m}^{-1}$ and a lifetime  ${\tau}_d{\sim}9$~ms.  The temperature of the BEC in the experiment was $T=21{\pm}20~{\rm nK}$, which belongs to the temperature regime shown in Figure \ref{fig2}.

The lifetimes ${\tau}_d=i/({\omega}^{(2)}q^2)$ of the fast and slow sound modes, obtained from Eq. (\ref{omega2f1}),  are plotted in Figure \ref{fig2} as a function of temperature for parameters applicable to the Steinhauer experiment. The dotted line corresponds to the slow modes with speed ${\pm}{\omega}^{(1)}_2$  and the solid line corresponds to the fast modes with speeds ${\pm}{\omega}^{(1)}_4$. The life times of the fast and slow sound modes cross at the same temperature at which the avoided crossing occurs in the fast and slow sound speeds (see Fig. 1).   The lifetime of the slow sound mode (dotted line) is of the order of microseconds above the crossing but drops to milliseconds below the crossing point.  The lifetime of the fast sound mode is of the order of milliseconds for temperatures above the crossing point, but then rapidly rises to microseconds for temperatures below the crossing point.  
The lifetime of the sound mode observed in the experiment was  ${\tau}_d{\sim}9$~ms. Thus we find, using the theory, that the temperature of the experiment was either $T = 11 {\pm} 1 {\rm nK}$ or $T = 67 {\pm} 5 {\rm nK}$.  

It should be noted that we use the  uniform density $n^{\rm eq}=9.71{\times}10^{19}~{\rm m}^{-3}$, while the density of a BEC in a trap varies slightly in the region that supports the sound wave, the change in the decay rates if the density were changed can be estimated to be 10$\%$.  The uncertainty in the prediction, due to a 10$\%$ uncertainty in the density is shown in Figure \ref{fig2} by the faint dashed lines that on either side of the result for
$n^{\rm eq}=9.71{\times}10^{19}~{\rm m}^{-3}$.  The uncertainty in the density does not significantly change the theoretical  prediction for the temperature at which sound waves in \cite{steinhauer12} was measured.
Thus, the value of the sound mode lifetime, predicted by \eqref{kineq1a} - \eqref{kineq1b} is consistent with that reported in \cite{steinhauer12}. 

\section{Conclusions}

A monatomic BEC has six hydrodynamic modes, two of which are transverse modes and describe viscous properties of the BEC, and the other four modes are longitudinal modes and describe sound mode propagation in the BEC.  A monatomic classical gas has one pair of propagating sound modes, one non-propagating thermal mode, and two non propagating viscous modes.  In a BEC, only the viscous modes are non-propagating. A dilute BEC has two pairs of sound modes, each of which is a mixture of density and temperature waves.  

The theory predicts that the two types of sound mode have different speeds and very different lifetimes. Using parameters from the Steinhauer experiment on a rubidium BEC, the theory indicates that one sound mode is long lived ($10^{-2}s$) and the other short-lived ($10^{-6}s$).  The identity of the long-live mode appears to switch at the temperature of the avoided crossing of the sound speeds.  At the temperature of the avoided crossing, neither mode lives a very long time. It has been suggested that this behavior of the sound modes could form the basis for an accurate way to determine the temperature of the BEC,   at  very low temperature.

\section{Acknowledgements}

Author LER thanks the Robert A. Welch Foundation (Grant No. F-1051) for support of this work. MBT was supported by  NSF Grant RNMS (Ki-Net) 1107444, ERC Advanced Grant DYCON.

\section{Appendix A}

We formulate below the form of the collision operator $\mathcal{G}[\eta]$

\begin{equation}
\begin{aligned}
\mathcal{G} [\eta]({{\bf k}_1}) \ =\ & -\N^{\rm eq}_{{\bf k}_1} (1+\N^{\rm eq}_{{{\bf k}_1})}~\Big(M({\bf k}_1)\eta({{\bf k}_1}) + \\
& \  \ \ +{\int}_{\bf{R}^3}d{\bf k}_2\frac{\N^{\rm eq}_{\kk_2}}{1+\N^{\rm eq}_{{{\bf k}_1}}}K({\bf k}_1,{\bf k}_2)\eta({{\bf k}_2})\Big)
\end{aligned}
\end{equation}
 with
\begin{equation}
\begin{aligned}
M({\bf k}_1) = & {\int}_{\bf{R}^3}d{\bf k}_2\frac{\N^{\rm eq}_{\kk_2}}{\N^{\rm eq}_{\kk_1}+1}\bigg{\{} 2A_0T_A(\kk_1,\kk_2)+A_0\frac{1+\N^{\rm eq}_{\kk_2}}{\N^{\rm eq}_{\kk_2}}T_B(\kk_1,\kk_2)~~~~~~~~~~~~\nonumber\\~
& +B_0Q_A(\kk_1,\kk_2)+B_0Q_B(\kk_1,\kk_2)+\frac{1}{3}\frac{1+\N^{\rm eq}_{\kk_2}}{\N^{\rm eq}_{\kk_2}}Q_C(\kk_1,\kk_2)\bigg{\}},~
\label{bigM1}\end{aligned}
\end{equation}
 \begin{equation}\begin{aligned}
K({\bf k}_1,{\bf k}_2)=&  \bigg{\{} 2A_0T_A({\bf k}_1,{\bf k}_2)-2A_0\frac{1+\N^{\rm eq}_{\kk_2}}{\N^{\rm eq}_{\kk_2}}T_B({\bf k}_1,{\bf k}_2) ~~~\nonumber\\
& -2A_0\frac{1+\N^{\rm eq}_{\kk_1}}{\N^{\rm eq}_{\kk_1}}T_B({\bf k}_2,{\bf k}_1)+B_0Q_A({\bf k}_1,{\bf k}_2) \nonumber\\
& -2B_0\frac{1+\N^{\rm eq}_{\kk_2}}{\N^{\rm eq}_{\kk_2}}R_A({\bf k}_1,{\bf k}_2)+2B_0Q_B({\bf k}_1,{\bf k}_2)~~~\nonumber\\
& -B_0\frac{1+\N^{\rm eq}_{\kk_2}}{\N^{\rm eq}_{\kk_2}}Q_C({\bf k}_1,{\bf k}_2)-B_0\frac{1+\N^{\rm eq}_{\kk_1}}{\N^{\rm eq}_{\kk_1}}Q_C({\bf k}_2,{\bf k}_1)\bigg{\}}.\end{aligned}
\end{equation}
The functions appearing in the above expressions are defined,
\begin{equation}\begin{aligned}
&A_0=\frac{4{\pi}N_0g^2}{(2\pi)^3{\hbar}V}, ~~~~
B_0=\frac{4{\pi}g^2}{(2\pi)^6{\hbar}}, ~~~~~~~~~~\\
T_A({\bf k}_1,{\bf k}_2)& = {\int}_{\bf{R}^3}d {\bf k}_3  \delta(1+2-3) (W^{12}_{1,2,3})^2  (\N^{\rm eq}_{{\bf k}_3 }+1), ~~~~~~~\\
T_B({\bf k}_1,{\bf k}_2)&={\int}_{\bf{R}^3}d {\bf k}_3   \delta(1-2-3) (W^{12}_{3,2,1})^2  (\N^{\rm eq}_{{\bf k}_3 }+1),~~~~~~~ \\
T_B({\bf k}_2,{\bf k}_1)&={\int}_{\bf{R}^3} d {\bf k}_3   \delta(2-1-3) (W^{12}_{3,1,2})^2  (\N^{\rm eq}_{{\bf k}_3 }+1),\\
Q_A({\bf k}_1,{\bf k}_2) &={\int}_{\bf{R}^3} d {\bf k}_3 d {\bf k}_4 \delta(1+2-3-4) (W^{22}_{1,2,3,4})^2  (\N^{\rm eq}_{{\bf k}_3 }+1)(\N^{\rm eq}_{{\bf k}_4}+1),\\
R_B({\bf k}_1,{\bf k}_2)&= {\int}_{\bf{R}^3}d {\bf k}_3  d {\bf k}_4  \delta(1+2-3-4) (W^{22}_{1,3,2,4})^2   \N^{\rm eq}_{{\bf k}_3 }(\N^{\rm eq}_{{\bf k}_4}+1),\\
Q_B({\bf k}_1,{\bf k}_2) &={\int}_{\bf{R}^3} d {\bf k}_3 d {\bf k}_4 \delta(1+2+3-4) (W^{31}_{4,3,2,1})^2   \N^{\rm eq}_{{\bf k}_3 }(\N^{\rm eq}_{{\bf k}_4}+1),\\
Q_C({\bf k}_1,{\bf k}_2) &= {\int}_{\bf{R}^3} d {\bf k}_3 d {\bf k}_4 \delta(1-2-3-4) (W^{31}_{1,2,3,4})^2   (\N^{\rm eq}_{{\bf k}_3 }+1)(\N^{\rm eq}_{{\bf k}_4}+1),\end{aligned}
\end{equation}
where $$ {\delta}(1+2-3-4){\equiv}{\delta}({\bf k}_1+{\bf k}_2-{\bf k}_3-{\bf k}_4){\delta}(E({\bf k}_1)+E({\bf k}_2)-E({\bf k}_3)-E({\bf k}_4)),$$ (with similar definitions for $\delta(1+2-3)$ and $\delta(1+2+3-4)$. etc), $B$ is the volume of the box of bosons under consideration, as explained in \eqref{hamexact1},
\begin{equation}
W^{12}_{1,2,3} = u_1 u_2 u_3 - u_1 v_2 u_3 - v_1 u_2 u_3 + u_1 v_2 v_3 + v_1 u_2 v_3 - v_1 v_2 v_3,
\end{equation}
\begin{equation}
W^{22}_{1,2,3,4} = u_1 u_2 u_3 u_4 + u_1 v_2 u_3 v_4 + u_1 v_2 v_3 u_4 + v_1 u_2 u_3 v_4 + v_1 u_2 v_3 u_4 + v_1 v_2 v_3 v_4
\end{equation}
and
\begin{equation}
W^{31}_{1,2,3,4} = u_1 u_2 u_3 v_4 + u_1 u_2 v_3 u_4 + u_1 v_2 u_3 u_4 + v_1 v_2 v_3 u_4 + v_1 v_2 u_3 v_4 + v_1 u_2 v_3 v_4.
\end{equation}
The Bogoliubov factors $u_i$ and $v_i$, $(i=1,2,3,4)$, are given by $$u_i=u_{k_1}=\frac{1}{\sqrt{2}}\sqrt{1+\frac{{\epsilon}_{{\kk}_i}+{\Lambda_0}}{E_{{\kk}_i}}}$$ and
$$v_i=v_{k_i}=\frac{1}{\sqrt{2}}\sqrt{\frac{{\epsilon}_{{\kk}_i}+{\Lambda_0}}{E_{{\kk}_i}}-1}.$$

\section{Appendix B}

When the BEC is in equilibrium, the mean field Hamiltonian (in the superfluid rest frame) takes the form \cite{ReichlGust:2013:TTF}
\begin{equation}
{\hat H}_0={\sum_i}\left[ ({\epsilon}_i-{\Lambda_0}){\hat a}_i^{\dagger}{\hat a}_i+\frac{\Lambda_0}{2}\left({\hat a}_i^{\dagger}{\hat a}_i+{\hat a}_i^{\dagger}{\hat a}_i\right)\right]=\frac{g}{2}N_0^2+{\sum_i}E_i{\hat b}_i^{\dagger}{\hat b}_i,
\end{equation}
where 
\begin{equation}
E_1=\sqrt{e_1^2-{\Lambda_0}^2} ~~{\rm with}~~e_1=\frac{{\hbar}^2k_1^2}{2m}+{\nu}^0-{\mu}=\frac{{\hbar}^2k_1^2}{2m}+{\Lambda_0}
\end{equation}
and we have used the Hugenholtz-Pines  relation ${\mu}={\nu}^0-{\Lambda_0}$ \cite{hugenholtz}. In terms of these equilibrium quantities, the Bogoliubov transformation parameters take the form
\begin{equation}
u_1=\frac{1}{\sqrt{2}}\sqrt{1+\frac{e_1}{E_1}},~~~v_1=\frac{1}{\sqrt{2}}\sqrt{\frac{e_1}{E_1}-1}
\end{equation}
Note also that 
\begin{equation}
u_1^2-v_1^2=1,~~~{\Lambda_0}(u_1^2+v_1^2)-2e_1u_1v_1=0,~~~e_1(u_1^2+v_1^2)-2{\Lambda_0}u_1v_1=E_1.
\end{equation}
This transformation has the property that
\begin{equation}
{\bar U}^{-1}_1{\cdot}
\left( \begin{array}{cc}
e_1 & {\Lambda_0}\\
- {\Lambda_0} &- e_1 \\
\end{array}\right)
{\cdot}{\bar U}_1={\bar U}_1{\cdot}
\left( \begin{array}{cc}
e_1 & -{\Lambda_0}\\
 {\Lambda_0} &- e_1 \\
\end{array}\right)
{\cdot}{\bar U}^{-1}_1 =\left( \begin{array}{cc}
E_1&0\\
0 &- E_1 \\
\end{array}\right).
\end{equation}
\section{Appendix C}

The collision operator operator ${\mathcal C}({\bf k},{\bf k}_1)$, that appears in Eq. (\ref{totkin2}), can be expanded in spherical harmonics which determine on the angular directions of the momenta ${\bf k}$ and ${\bf k}_1$. 
\begin{equation}
{\mathcal C}({\bf k},{\bf k}_1) ={\sum_{\ell=0}^{\infty}}{\sum_{m=-\ell}^{\ell}}{\mathcal C}_{\ell}(k,k_1) {\rm Y}_{\ell}^m({\hat {\bf k}}){\rm Y}_{\ell}^{m*}({\hat {\bf k}_1})
\label{coll1}
\end{equation}
The collision operator ${\mathcal C}({\bf k},{\bf k}_1)$ is a symmetric operator and has a complete set of orthonormal eigenfunctions. 

The eigenvalues
 ${\lambda}_{\beta,\ell}$ and eigenstates ${\Upsilon}_{\beta,\ell,m}({\bf k}_1)$  of the operator
 ${\mathcal C}({\bf k}_1,{\bf k}_2)$, satisfy the conditions
\begin{eqnarray}
{\int}~d{\bf k}_2~C({\bf k}_1,{\bf k}_2) {\Upsilon}^{(0)}_{\beta,\ell,m}({\bf k}_2)={\lambda}_{\beta,\ell} {\Upsilon}^{(0)}_{\beta,\ell,m}({\bf k}_1).
\end{eqnarray}
and
\begin{eqnarray}
{\int}~d{\bf k}_1~{\Upsilon}^{(0)}_{\beta,\ell,m}({\bf k}_1)~C({\bf k}_1,{\bf k}_2) ={\lambda}_{\beta,\ell} {\Upsilon}^{(0)}_{\beta,\ell,m}({\bf k}_2).
\end{eqnarray}
The eigenvalues  ${\lambda}_{\beta,\ell}$ are independent of $m$ due to the angular symmetry of the collision operator.

The bogolon momentum and energy are conserved during collisions, although bogolon number is not. Therefore, $\mathcal{G}_{{\bf k}_1} \{ h \}$, acting on four conserved quantities, $h=E_k$, $h=k_x$, $h=k_y$, and $h=k_z$, gives zero. We can use this fact to form four eigenstates of  $C({\bf k}_1,{\bf k}_2)$. We write them in the form
\begin{eqnarray}
{\Upsilon^{(0)}_{0,0,0}({\bf k}_1)={\Upsilon}_{0,0}(k_1){\rm Y}_0^0({\hat {\bf k}}_1),~~~
\Upsilon}^{(0)}_{0,1,0}({\bf k}_1)={\Upsilon}_{0,1}(k_1){\rm Y}_1^0({\hat {\bf k}}_1), ~~\nonumber\\
{\Upsilon}^{(0)}_{0,1,1}({\bf k}_1)={\Upsilon}_{0,1}(k_1){\rm Y}_1^1({\hat {\bf k}}_1),  ~~~
{\Upsilon}^{(0)}_{0,1,-1}({\bf k}_1)={\Upsilon}_{0,1}(k_1){\rm Y}_1^{-1}({\hat {\bf k}}_1).
\end{eqnarray}
where
${\Upsilon}_{0,0}(k)=D_{0,0}E_k\sqrt{k^2\N_k^{\rm eq}\F_k^{\rm eq}}$ and  ${\Upsilon}_{0,1}(k)=D_{0,1}k\sqrt{k^2\N_k^{\rm eq}\F_k^{\rm eq}}$. The quantities $D_{\beta,\ell}$,  are normalization constants given by
\begin{eqnarray}
D_{0,0}=\left( {\int_0^{\infty}}dk k^2E_k^2\N_k^{\rm eq}\F_k^{\rm eq} \right)^{-1/2}~~{\rm and}~~D_{0,1}=\left( {\int_0^{\infty}}dk k^4\N_k^{\rm eq}\F_k^{\rm eq} \right)^{-1/2}.
\label{CYseries1}
\end{eqnarray}
The corresponding eigenvalues
${\lambda}_{\beta,\ell}$  are independent of $m$  and degenerate so that  ${\lambda}_{0,0}={\lambda}_{0,1}=0$ and ${\lambda}_{0,1}$ is three-fold degenerate.
The eigenstates can be orthonormalized so that
\begin{equation}
{\int_0^{\infty}}dk_1{\int}d{\Omega}_1~{\Upsilon}^{(0) *}_{\beta_1,\ell_1,m_1}({\bf k}_1){\Upsilon}^{(0)}_{\beta_2,\ell_2,m_2}({\bf k}_1)={\delta}_{\beta_1,\beta_2}{\delta}_{\ell_1,\ell_2}{\delta}_{m_1,m_2}
\end{equation}
and ${\int_0^{\infty}}dk_1~{\Upsilon}^*_{\beta_1,\ell}(k_1){\Upsilon}_{\beta_2,\ell}(k_1)={\delta}_{\beta_1,\beta_2}$.  We can now write the spectral decomposition of the collision operator
\begin{eqnarray}
C({\bf k}_1,{\bf k}_2)={\sum_{\beta=0}^{\infty}}{\sum_{\ell=0}^{\infty}}{\sum_{m=-\ell}^{\ell}}~{\lambda}_{\beta,\ell}~{\Upsilon}^{(0)}_{\beta,\ell,m}({\bf k}_1){\Upsilon}^{(0) *}_{\beta,\ell,m}({\bf k}_2) \nonumber\\
\label{CYseries1}
\end{eqnarray}
where ${\mathcal C}_{\ell}(k_1,k_2)={\sum_{\beta=0}^{\infty}}~{\lambda}_{\beta,\ell}~{\Upsilon}_{\beta,\ell}(k_1){\Upsilon}^*_{\beta,\ell}(k_2)~$.
These four eigenfunctions form the basis for the hydrodynamic modes in the BEC.  They correspond to quantities that are conserved on the microscopic scale.

\pagebreak

~~~~LIST OF FIGURES
\vspace{0.1cm}

\begin{figure}[h]
\includegraphics[width=0.85\columnwidth]{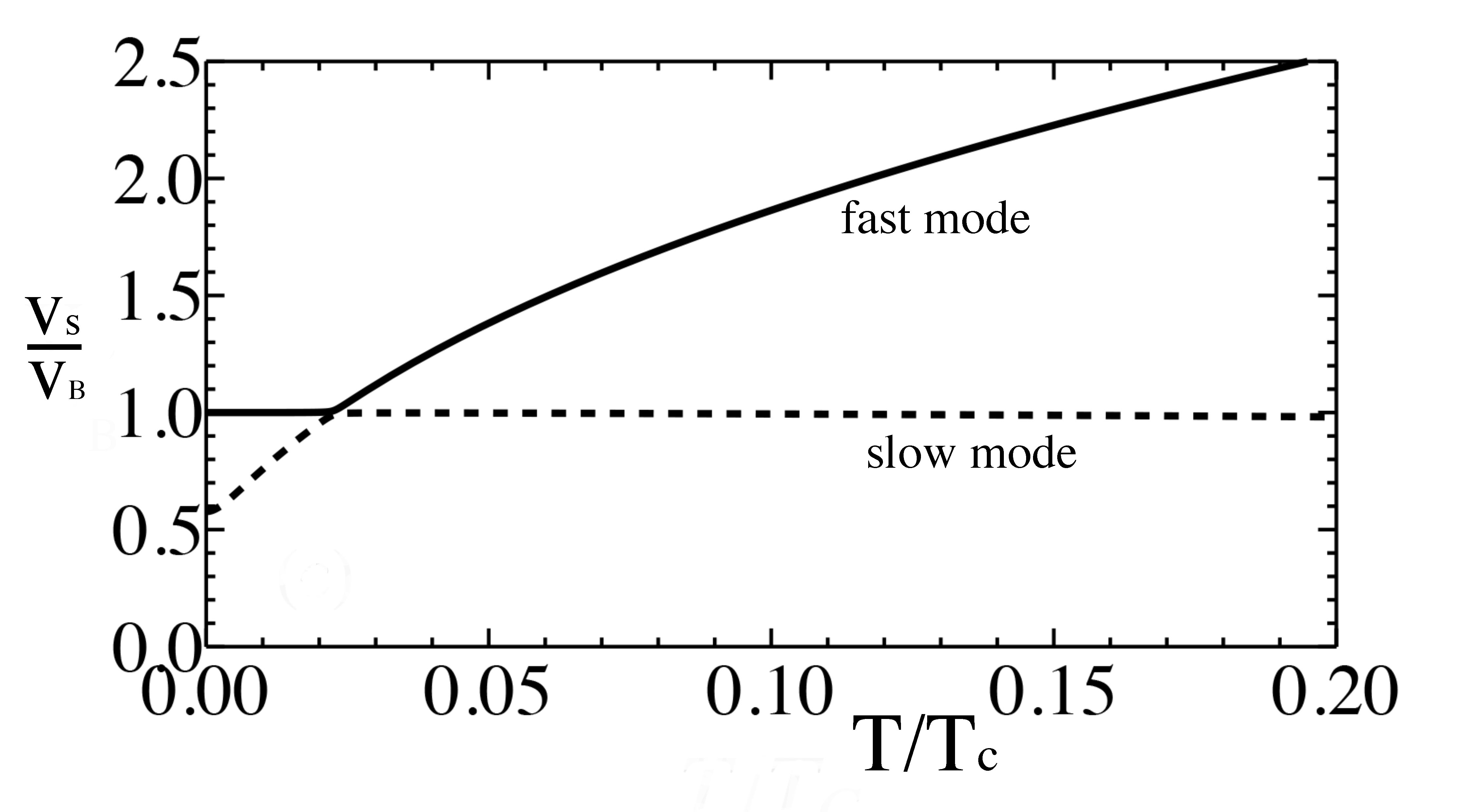}
\caption{Propagation speeds of the fast (solid) and slow (dashed) longitudinal modes at $na^3 = 10^{-6}$, in units of $v_B$ [based on \cite{ ReichlGust:2014:DOH}].
\label{fig1}}
\end{figure}

\begin{figure}[h]
\includegraphics[width=0.85\columnwidth]{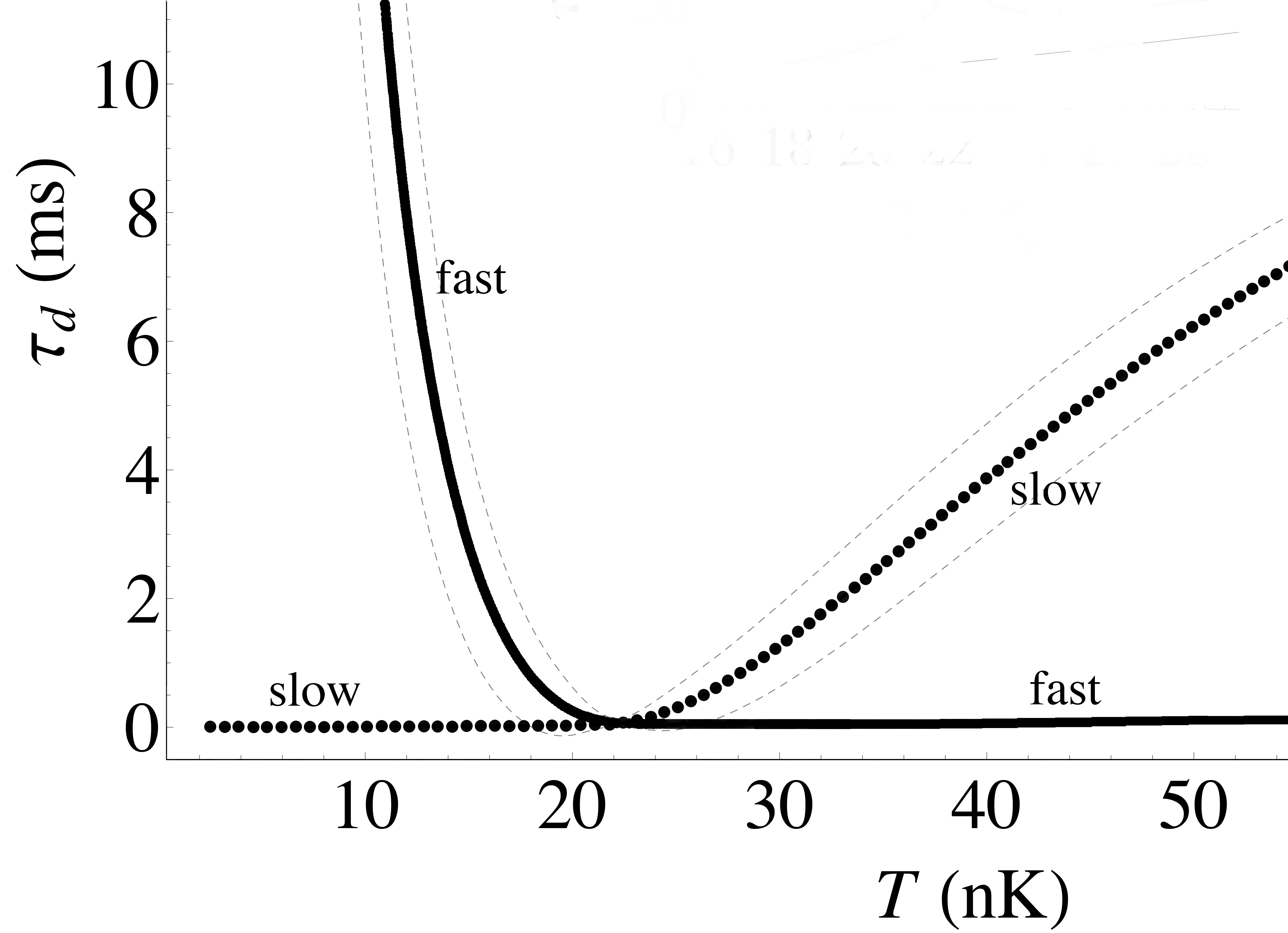}
\caption{Decay times $\tau_d$ for $^{87} {\rm Rb}$ at $q= 0.35~ \mu {\rm m}^{-1}$ in the fast (solid) sound mode  and slow (dotted) sound modes. The faint lines give uncertainty in the results. The experimental value of $\tau_d=9$ ms was obtained for a temperature of $T = 21 {\pm}20 {\rm nK}$.  [based on \cite{ ReichlGust:2014:DOH}].
\label{fig2}}
\end{figure}

\end{document}